\def\showcomments{}
\ifdefined\drafting
	\documentclass[aps,prb,notitlepage,preprint,floatfix,tightenlines]{revtex4-2}
\else
	\documentclass[aps,prb,notitlepage,reprint]{revtex4-2}
	
\fi

\ifdefined\drafting
	\usepackage[a4paper,left=0.25in,right=1.8in,head=21.0pt]{geometry}
\else
	
\fi

\ifdefined\showcomments{}
    \usepackage[commentmarkup=uwave,todonotes={textsize=scriptsize,textwidth=1.6in}]{changes}
	\makeatletter \def\@captype{figure} \makeatother 
\setcommentmarkup{{\IfIsColored{\color{authorcolor}}{}[#1]\textsuperscript{#2\arabic{authorcommentcount}}}}
\else
	\usepackage[final]{changes}
\fi

\newcommand{\markhigh}[1]{\bgroup\markoverwith				
  {\textcolor{#1}{\rule[-.5ex]{2pt}{2.5ex}}}\ULon}

\ifdefined\removefig
	\usepackage{comment} 	
	\excludecomment{figure}	
		
\else
\fi

\usepackage{braket}
\usepackage{amsmath}
\usepackage{amssymb}
\usepackage{mathtools}
\usepackage{siunitx}
\usepackage{graphicx}
\usepackage{nameref}
\usepackage[colorlinks=true,linkcolor=blue]{hyperref}
\usepackage[capitalise]{cleveref}
\usepackage{bm}
\usepackage{environ}
\usepackage[percent]{overpic} 
\newcommand{\ii}{\mathrm{i}}
\newcommand{\ee}{\mathrm{e}}

\NewEnviron{ignore}{} 

\definechangesauthor[name=Slava Kashcheyevs, color=blue]{VK}
\definechangesauthor[name=Girts Barinovs, color=brown]{GB}
\definechangesauthor[name=Agris Buzs, color=teal]{AB}

\begin{document}
\title{Long-lived Laughlin pairs in a depleted quantum Hall edge channel}

\author{Girts Barinovs}
\affiliation{Department of Physics, Faculty of Science and Technology, University of Latvia, Riga LV-1004, Latvia}
\author{Agris Buzs}
\affiliation{Department of Physics, Faculty of Science and Technology, University of Latvia, Riga LV-1004, Latvia}
\author{Vyacheslavs Kashcheyevs}
\affiliation{Department of Physics, Faculty of Science and Technology, University of Latvia, Riga LV-1004, Latvia}

\begin{abstract}
On-demand sources and mesoscopic beam splitters now allow individual ballistic
electrons to collide in depleted quantum Hall edge channels, where their
unscreened Coulomb interaction acts as a strong, controllable nonlinearity.
Theory suggests a more striking possibility: in a strong magnetic field, the
same repulsion can drive quantized relative circulation, allowing two electrons
to propagate together as a positive-energy Laughlin pair. The relevance of such pairs to
experiment, however, depends on quantitative lifetimes in realistic guiding
potentials and on whether the proposed collision pathway to pair formation
survives full two-dimensional dynamics. Here we develop a microscopic theory
of quasibound Laughlin pairs using the physical two-electron Hamiltonian. For a
general local electric-field gradient, we determine the dissociation threshold, number of
quasibound states, and decay rates. Complex scaling and analytic tunneling
theory show that lifetimes grow exponentially with pair energy above threshold.
Applied to reported GaAs parameters, the theory indicates that existing devices
may already support the lowest spin-polarized pair, 
with a conservative  lifetime
estimate roughly two orders of magnitude longer than typical propagation times.
We also simulate a collision with the full finite-field Hamiltonian,
providing both a framework for nonlinear two-electron quantum dynamics and
evidence for Laughlin-pair formation in a representative two-electron
collision.  We use Husimi
distributions and their zeros to visualize both quasibound resonances and
transient collision states in phase space. Together, these results
place the preparation, propagation, and detection of repulsively paired
electrons within reach of existing single-electron circuit technology. They
also identify kinematic stabilization under constrained one-dimensional
propagation as a pairing mechanism that may extend to anyonic quantum Hall edge
excitations.
\end{abstract}
\maketitle

\section{Introduction}

Electron quantum optics seeks to generate, manipulate, and detect elementary
electronic excitations one by one~\cite{Bocquillon2014,Bauerle2018}. In quantum
Hall edge channels, on-demand sources~\cite{Feve2007} and gate-controlled beam
splitters have enabled the partitioning of individual
electrons~\cite{Bocquillon2012}, two-electron
interference~\cite{Bocquillon2013}, and quantum tomography of electrical
currents~\cite{Bisognin2019}. Much existing work concerns excitations
propagating on top of a Fermi sea~\cite{Feve2007,Dubois2013}, for which the
optical analogy remains productive even when Coulomb coupling to other edge
channels causes relaxation and charge fractionalization~\cite{Freulon2015}.

A complementary regime is accessed by dynamic-quantum-dot
sources~\cite{Pekola2013,ROPP2015}, which emit on-demand wave packets at
tuneable energies into channels locally depleted of other
carriers~\cite{kaestner2010d,Fletcher2013,Hermelin2011,McNeil2011,Kataoka2016a,Freise2019}.
The emitted electrons are separated from the Fermi sea in energy and space.
In collisions at smooth quantum point contacts, their partitioning is governed
by direct Coulomb
repulsion~\cite{Pavlovska2023,Ubbelohde2023,Fletcher2023NatNa..18..727F,Wang2022}.
Solitary-electron circuits therefore give controlled access to few-electron
dynamics in the strong-interaction regime~\cite{Shaju2025}. They also pose a
question that has no direct optical counterpart: can repulsion itself create a
propagating two-electron composite?

The origin and stability of such a state can be described within a simple
intuitive picture. In a strong magnetic field, the fast cyclotron motion of a
charged particle is quantized (Landau quantization) and the drift of the guiding
center is always perpendicular to the applied force. For on-demand electronic
excitations in two-dimensional heterostructures, this guiding force is supplied
by a non-uniform in-plane electric potential $\phi(x,y)$, controlled by the
geometry of the sample edges and the field effect of capacitive gates patterned
on top. 
With  \(\bm E=-\nabla\phi\) and 
\(\bm B=B_z\hat{\bm z}\) fields,  $\bm E\times\bm B$ drift follows equipotential lines, and its
velocity $\bm v_{\rm d}=\{-\partial_y\phi,\partial_x\phi\}/B_z$ is independent
of the charge of the excitation.

\begin{figure*}
    \centering
    \begin{overpic}[width=0.35\textwidth]
      {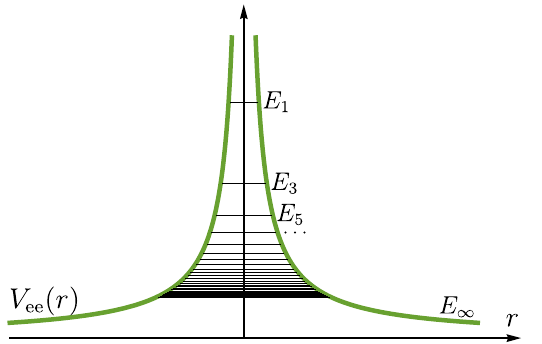}\put(2,60){\footnotesize\textbf{(a)}}
    \end{overpic}
    \hspace{0.02\textwidth}
    \begin{overpic}[width=0.35\textwidth]
      {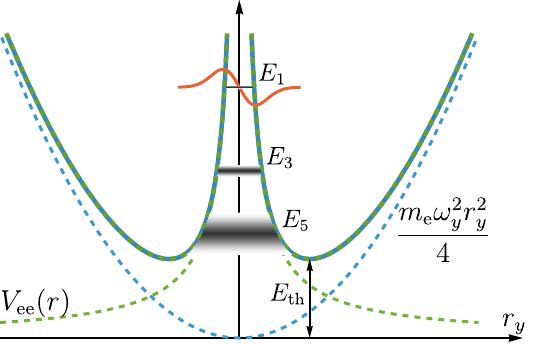}\put(2,60){\footnotesize\textbf{(b)}}
    \end{overpic}
    \caption{
    Magnetic-field binding of Laughlin pairs and destabilization by electric-field gradient.
    (a) In a flat external potential, relative drift motion of two electrons in their  lowest-Landau-level state  is quantized and samples the
    repulsive Coulomb interaction $V_{\text{ee}}(r)$ (green curve) at discrete energy levels $E_m$. The odd-$m$ sequence relevant to spin-polarized electrons is shown;
    its positive energies decrease towards the separated-electron limit as the
    pair radius $r$ grows. (b) Schematic cut of the relative-motion potential along
    the stable principal axis, $r_x=0$, when an external electric-field gradient is
    present. The Coulomb and quadratic contributions as functions of $r_y$ are shown by dashed
    curves. Their interaction-induced stationary points are saddles of the full
    two-dimensional landscape; the transverse open direction ($r_x$) responsible for
    dissociation is not visible in this cut. Only a finite number of quasibound
    states lie above the critical energy $E_{\rm th}$, and their lifetimes grow
    exponentially with $E_m-E_{\rm th}$.}
\label{fig:mainSketch}
\end{figure*}

When the external guiding potential is absent (uniform $\phi$), all
single-electron states within one Landau level are degenerate and the motion is
quenched, but this changes for two nearby electrons. Their Coulomb repulsion
generates motion tangential to the central force, resulting in circular
relative orbits around the center of mass. The angular momentum of this motion
is quantized to $m\hbar$. These relative-motion states were first considered by
Laughlin~\cite{Laughlin1982} as a stepping stone to his theory of the
fractional quantum Hall effect~\cite{Laughlin1983}, hence the name
\emph{Laughlin pairs}. Their repulsive origin is reflected in an inverted
spectrum: the positive pair energy $E_m$ decreases with $m$ as the orbit grows.
For the Coulomb potential $V_{\rm ee}(r)\propto 1/r$, $E_m\propto m^{-1/2}$ at
large $m$, as illustrated in Fig.~\ref{fig:mainSketch}(a).

Here we consider the local conditions for stability of a Laughlin pair in a
smooth external electric potential. In contrast to an exciton or positronium,
which consists of opposite charges, a uniform electric field cannot break a
Laughlin pair: both electrons acquire
an identical shift in drift velocity. Stability is therefore
determined by the next order in the expansion of $\phi$, the electric-field
gradient tensor $\hat V$, or equivalently the Hessian of $\phi$ at the pair
location. The decay products are two asymptotically separated electrons, each
of charge $-e$ and subject to the same external electrostatic landscape as the
pair of total charge $-2e$. To carry away the positive internal energy $E_m$,
their external potential energy must grow with separation in at least one
direction, requiring a positive eigenvalue of $-e\hat V$. Open outgoing
trajectories require the other eigenvalue to be non-positive at quadratic
order. Together these conditions imply a finite quadrupole anisotropy, which
couples states whose angular momenta differ by $\pm2$ and provides the matrix
elements for pair decay. As sketched in Fig.~\ref{fig:mainSketch}(b), the bulk
Laughlin states then evolve into a finite set of quasibound resonances whose
lifetimes depend strongly on their distance from the dissociation threshold.

Reports of correlated multiparticle transport in populated quantum Hall
edges~\cite{Choi2015Pairing,Biswas2023Pairing,Yang2024PairsTriplets,Ghosh2025Bunching}
provide a broader experimental context. At integer filling, interference and
noise measurements have been interpreted as signatures of electron
pairing~\cite{Choi2015Pairing,Biswas2023Pairing}; graphene interferometry has
revealed correlations assigned to electron pairs or
triplets~\cite{Yang2024PairsTriplets}; and a recent fractional-edge experiment
has been interpreted as coherent bunching and dissociation of several
anyons~\cite{Ghosh2025Bunching}. These experiments motivate the broader
question of propagating electronic composites, but their relation to the
Laughlin pairs studied here remains open. The depleted-channel setting instead
defines a distinct few-electron problem in which the constituents, internal
wave function, dissociation channel, and lifetime can be treated
microscopically.

Previous work established how Coulomb interactions shape collisions of two
chiral electrons at smooth quantum point
contacts~\cite{Pavlovska2023,Ubbelohde2023,Fletcher2023NatNa..18..727F}, while
Ref.~\onlinecite{Silvestrov2025} extended this framework to the guided motion
and tunneling-assisted formation of Laughlin pairs (``electron molecules''),
including parity-selective survival as a possible route to heralded,
spatially distributed spin entanglement. The present work develops the
microscopic stationary theory of the pair resonances in the physical
two-electron Hamiltonian. We focus on odd relative parity, appropriate to
spin-polarized electrons, and approximate the smooth landscape locally to
quadratic order. After projection to the lowest Landau level (LLL), we determine the
curvature-controlled existence threshold and the number of supported pair
branches, calculate their complex energies and decay widths, and derive
analytic approximations for the rapid suppression of pair decay rates as
$E_m-E_{\rm th}$ increases.

The strong magnetic fields that stabilize Laughlin pairs also suppress
Landau-level mixing, making the lowest-Landau-level projection increasingly
accurate for GaAs~\cite{Laughlin1982,girvin1984formalism}. The resulting
dimensionless formulation combines the material parameters, magnetic field, and two principal
landscape curvatures into two dimensionless parameters, allowing the number of
supported branches and their lifetimes to be estimated across smooth
landscapes. Applied to parameters reported for GaAs chiral single-electron
propagation and collision
circuits~\cite{Kataoka2016a,Ubbelohde2023,Fletcher2023NatNa..18..727F,Freise2019},
this framework places existing devices in the relevant regime and provides
quantitative guidance for the design of future experiments to prepare,
transport, and detect Laughlin pairs.

To expose the internal structure of these resonances, we connect the Wigner
phase-space language familiar from electron quantum
optics~\cite{Ferraro2013} to the Husimi
density~\cite{CahillGlauber1969DensityOperators}, which in the lowest Landau
level is also the physical relative-coordinate density, and to the
complementary stellar representation in terms of wave-function
zeros~\cite{Chaubaud2020}. These representations show how an ideal Laughlin
state deforms and acquires outgoing scattering structure while retaining a
recognizable branch identity through the $m$ internal zeros inherited from the
rotationally symmetric state. We then test this stationary description by propagating one representative
collision under the full relative-motion Hamiltonian at finite magnetic field,
without lowest-Landau-level projection. Once the prompt scattering components
have escaped, the evolution isolates a long-lived core whose density and
central zero agree qualitatively with the independently calculated $m=1$
resonance and whose decay rate is consistent after the finite-field energy
correction. 
From the inferred weight of the exponentially decaying quasistationary state at the onset of the exponential regime, we obtain an operational
  Laughlin-pair formation yield above $5\%$ for this collision.

Recent theoretical studies 
explored microscopic conditions under which 
fractional quasiparticles in the
bulk of fractional quantum Hall phases may form energetically bound
molecules~\cite{GattuJain2025MolecularAnyons,Xu2025AnyonClusters,Li2026BoundAnyons,WangZaletel2026AnyonMolecules}.
The pairing mechanism studied here instead raises a distinct question for populated
quantum Hall edges: whether the reduced decay-product phase space of a
one-dimensional chiral channel may stabilize, over a finite propagation
length, a topologically allowed internal anyon state even if it is not energetically
favorable relative to separated anyons; we discuss this kinematic
possibility in Section~\ref{sec:implicationsOutlook}.

The paper is organized as follows. Section~\ref{sec:model} introduces the
two-electron Hamiltonian, its lowest-Landau-level projection, and the
Wigner--Husimi--wave-function-zero dictionary. Section~\ref{sec:stabilityAndRates} describes
energies and decay rates of quasibound Laughlin pairs, and
Section~\ref{sec:resVis} visualizes the corresponding resonant wave functions.
Section~\ref{sec:pairFormation} presents a finite-field calculation of a representative two-electron collision, 
identifies the metastable $m=1$ remnant
from its structure and decay, and gives an operational
estimate for Laughlin-pair formation yield in that collision.
Section~\ref{sec:implicationsOutlook} relates the stability results of Section~\ref{sec:stabilityAndRates}  to existing
devices and discusses potential implications for anyon physics, before
Section~\ref{sec:conclusions} concludes.

\section{Model\label{sec:model} and methods}

\subsection{Notation and characteristic energies\label{sec:notaitonscales}}
The characteristic Coulomb energy of Laughlin pairs is
$E_\mathrm{C}=\hbar (\omega_\mathrm{c} \omega^{\star})^{1/2}$
where the cyclotron frequency $\omega_\mathrm{c}= eB/m_{\mathrm{e}}$ is determined by the 
magnetic-field magnitude $B$  and the single-electron effective mass $m_{\text{e}}$
and  the effective Hartree $\hbar \omega^{\star}=
m_{\mathrm{e}}  e^4/(4\pi\epsilon_0\epsilon \hbar )^2$ characterizes the strength of the Coulomb potential with relative dielectric constant $\epsilon$,
\begin{align} \label{eq:pureCoulomb}
    V_\mathrm{ee}(r) = \frac{e^2}{4 \pi \epsilon_0 \epsilon} 
    \frac{1}{r} \, .
\end{align}
For GaAs with $\epsilon \approx 12$, we use $\hbar \omega^{\star}=\SI{12}{meV}$ which gives $E_\mathrm{C} \approx \SI{15}{meV}$ at $B=\SI{10}{T}$.  
Note that $E_\mathrm{C}$ sets the scale for the excitation gap  in fractional quantum Hall states of Laughlin sequence with filling factors $1/m$, and the pairs can be seen as the minimal ``molecules'' of the corresponding quantum liquids.

In the large magnetic field limit,   the energy of an isolated pair with angular momentum $m \hbar$ (with respect to fully decoupled two electrons at $m \to \infty$) is
$E_m= E_\mathrm{C} \varepsilon_m^{(0)}$ where the dimensionless unperturbed energies for the interaction potential \eqref{eq:pureCoulomb}
are \cite{Laughlin1982}
\begin{align} \label{eq:pairs}
    \varepsilon_m^{(0)} = \frac{\sqrt{\pi}}{2} \frac{(2m)!}{(2^m m!)^2} =\frac{1}{2 \sqrt{m}} +O(m^{-3/2}) \, .
\end{align}
We consider Laughlin pairs in  an external smooth electric potential  $\phi(\bm{r})$ with a constant electric field gradient tensor $\hat{V}$, characterized by its principal values  $\omega^2_y = -(e/m_\text{e}) \partial_y^2\phi >0$ 
and $\omega^2_x =-(e/m_\text{e}) \partial_x^2\phi$.
The Cartesian coordinates $\bm{r}=(x,y)$ are aligned with the principal axes of $\hat{V}$ and
we take \(\omega_y^2>\omega_x^2\) and use its principal direction as
the \(y\) axis.
 Unbound trajectories for pair decay products correspond to $\omega_x^2 \leq 0$, and we denote $\kappa^2= -\omega_x^2/\omega_y^2$ to characterize anisotropy ($\kappa^2=1$ corresponds to traceless $\hat{V}$ and pure quadrupole anisotropy of the field external to the pair).

The dissociation threshold for pairs due to electric field gradient is determined by the competition between the internal and the external electric forces, and as such is independent of the quantum effects and the magnetic field. This competition can be quantified by the distance $d_0$ and the energy scale 
$U=V_\mathrm{ee}(d_0)$ at which the two are balanced (interaction energy at the transition state) in the direction of strongest confinement, $y$~\cite{Pavlovska2023}.
For the Coulomb potential $V_\mathrm{ee}(r) = U d_0/r$ and the quadratic $\phi(\bm{r})$ we have  $U=\hbar(\omega^{\star} \omega_y^2/2)^{1/3}$.

If the magnetic field is strong enough, $E_\mathrm{C} > U$, there will be long-lived pairs.
In the following we use a dimensionless ratio $\chi =(U/E_\mathrm{C})^{3}=\omega_y^2/(4  \omega_\mathrm{c}^{3} \omega^{\star} )^{1/2}$ as a parameter of the dimensionless Hamiltonian and  the scaled stability diagram.

\subsection{Hamiltonian\label{sec:Hamiltonian}}

The full Hamiltonian of two interacting electrons in two dimensions is 
\begin{align}
\mathcal{H}_{\text{tot}}=\mathcal{H}_{1\text{e}}(\bm{r}_1)+\mathcal{H}_{1\text{e}}(\bm{r}_2) + V_{\text{ee}}( \left\|\bm{r}_2-\bm{r}_1\right\|) 
 \, , \label{eq:oneeletronham}
\end{align}
where 
\begin{align} \label{eq:Hsp}
  \mathcal{H}_{1\text{e}}(\bm{r}_j)= \frac{1}{2 m_{\text{e}}}
    \left [- \ii \hbar \nabla_{\bm{r}_j} + e \mathbf{A}(\bm{r}_j) \right 
    ]^2 - e \phi(\bm{r}_j) \, , 
    \end{align}
and the vector potential in the symmetric gauge, $\bm{A}(\bm{r}_j)  = \bm{B} \times \bm{r}_j /2$, describes a uniform out-of-plane magnetic field, $\bm{B}= B_z \hat{z} = -B\,\hat{\bm{z}}$ (with $B>0$ this sets our chirality sign convention, used in \cite{Pavlovska2023}).
We neglect spin-orbit interaction and suppress spin indices.

We approximate the smooth electric potential  $\phi(\bm{r})$ in the relevant region around the center of mass 
 ${\bm{R}=(\bm{r}_1+\bm{r}_2)/2}$ up to quadratic order with electric field gradient tensor parameters $\omega_{y}^2$ and $\omega_{x}^2$ as defined above. With these assumptions,
 the variables in Eq.~\eqref{eq:oneeletronham} separate as 
$\mathcal{H}_{\text{tot}} =
 \mathcal{H}_{\bm{R}} +
   \mathcal{H}_{\bm{r}}$
with
\begin{align}  \label{eq:HrR}
  \mathcal{H}_{\bm{R}}= 
  &\frac{1}{4 \, m_\text{e}}
    \left [- \ii \hbar \nabla_{\bm{R}} + 2 e \mathbf{A}(\bm{R}) \right 
    ]^2 -2 e \phi (\bm{R}) \, . 
\end{align}
The motion of the pair as a whole corresponds to that of a particle with double the charge and double the mass of a single electron, following the level lines of $ \phi (\bm{R})$ with the same drift velocity $\bm{v}_\mathrm{d}$ as a single electron. 

The relative coordinate $\bm{r}=\bm{r}_2-\bm{r}_1$ is governed by 
\begin{align} \label{eq:Hr}
  \mathcal{H}_{\bm{r}}= & \frac{1}{ m_{\text{e}}}
    \left [- \ii \hbar \nabla_{\bm{r}} + (e/2) \mathbf{A}(\bm{r}) \right]^2+  \nonumber \\ &
    +\frac{m_\text{e}\omega_x^2 r_x^2}{4}
    +\frac{m_\text{e}\omega_y^2 r_y^2}{4} +  V_{\text{ee}}( \left\|\bm{r}\right\|)  \, ,
\end{align}
where $(r_x,r_y)$ are components of $\bm{r}$ along the principal axes of the electric field gradient tensor. Equation \eqref{eq:Hr}  corresponds to a fictitious particle with reduced mass $m_{\text{e}}/2$ and charge $-e/2$, 
moving in a repulsive central potential 
and an electric field gradient.
The potential energy part of Eq.~\eqref{eq:Hr} along $y$  axis is illustrated in Fig.~\ref{fig:mainSketch}(b).

Exchange symmetry of two-electron  states in the absence of spin-orbit coupling separates the Hilbert space into sectors of well-defined total spin (either singlet or triplet), with the orbital part being an eigenstate of parity transformation $\bm{r} \to -\bm{r}$ (odd for spin triplet, even for spin singlet).
Keeping in mind scenarios in which individual electrons and pairs are sourced from a spin-polarized  quantum Hall edge state, we will focus on wave functions that are odd in relative coordinate $\bm{r}$.

The Zeeman energy contribution 
(not included in the Hamiltonian \eqref{eq:oneeletronham}) is 
\begin{equation}
    E_\mathrm{Z}(M)=-\hbar \omega_\mathrm{c}\frac{g}{2}\frac{m_\mathrm{e}}{m^{(\text{vac})}_\mathrm{e}}M \, ,
\end{equation}
where $g$ is the appropriate Landé factor, ${m^{(\text{vac})}_\mathrm{e}}$ is the free electron mass  in the vacuum, and $M$ is the total magnetic spin quantum number ($M=\pm 1, 0$ for the three triplet components).

\subsection{Projection to the LLL and dimensionless form\label{sec:proj}}

Projection of a single-particle wave function $\psi(\bm{r})$ to  the LLL~\cite{girvin1984formalism} corresponds to representing  it as  \begin{align} \label{eq:psiLLL}
\psi(\bm{r}) = l^{-1} \sum_m c_m \varphi_m(r_x/l + \ii r_y/l)  + \psi^{\perp}(\bm{r}) \, ,
\end{align}
where the basis functions, 
\begin{align} \label{eq:1eLaughlin}
 \varphi_m(z) = \frac{z^m}{\sqrt{2^{m+1} m! \pi}} \ee^{-|z|^2/4}  \, ,
\end{align}
are the eigenstates of $\mathcal{H}_{1 \text{e}}(\bm{r})$ in uniform electric potential with degenerate eigenvalues $\hbar \omega_\mathrm{c}/2$;  $\psi^{\perp}(\bm{r})$ formally accounts for amplitudes in higher Landau levels.
In applying Eq.~\eqref{eq:1eLaughlin} to one electron, $\bm{r} \to \bm{r}_j$, $l$ is the standard magnetic length, while for  
 the relative motion the appropriate value of $l$ is $\sqrt{2}$ times larger, 
$l=\sqrt{2 \hbar/ (m_{\text{e}} \omega_\mathrm{c})}$, because of the reduced mass.  

Projection of the interaction potential \eqref{eq:pureCoulomb} alone gives directly the energies $E_m^{(0)}$ of the unperturbed pairs, Eq.~\eqref{eq:pairs}. This corresponds to Haldane pseudopotential representation of electron-electron interaction in the LLL~{\cite{Haldane1983}. Using $E_\mathrm{C}$ as the unit of energy and $l$ as unit of length, the
projection of $\mathcal{H}_{\bm{r}}$ onto the LLL can be  written as $\mathrm{H} =(\mathcal{P} \mathcal{H}_{\rm{r}} \mathcal{P}^{\dagger}- E_{\text{LL0}}\, \mathcal{P}\mathcal{P}^{\dagger} )/E_\mathrm{C}$ with  
\begin{align} 
 \label{eq:projected}
 \mathrm{H} & = \sum_m  \varepsilon_m^{(0)} \ket{m}\bra{m} + 
    \chi \left(\hat{\rm{y}}^2-\kappa^2 
  \hat{\rm{x}}^2 \right) \, ,
  \end{align}
  where  $\varepsilon_m^{(0)}$ are unperturbed Laughlin energies given by Eq.~\eqref{eq:pairs}, 
  $\ket{\psi} = \mathcal{P} \psi(\bm{r})=\sum_m c_m \ket{m}$ projects an arbitrary wave function onto a reduced dimensionality state vector,  and $\mathcal{P} \psi^{\perp}(\bm{r})=0$, cf.~Eq.~\eqref{eq:psiLLL}. 
  The basis of Laughlin pair states with angular momentum $m \hbar$ projected onto the LLL [Eq.~\eqref{eq:1eLaughlin}] 
   forms the standard Fock basis $\ket{m}$ of a harmonic oscillator,  $(\hat{\mathrm{x}}^2+ \hat{\mathrm{y}}^2) \ket{m} = (2 m+1) \ket{m}$. Here the non-commuting guiding center coordinates \cite{girvin1984formalism} are defined by 
   the projections of $\bm{r}$  onto the LLL, $(\hat{\mathrm{x}},\hat{\mathrm{y}})=(\mathcal{P} r_x \mathcal{P}^{\dagger},\mathcal{P} r_y \mathcal{P}^{\dagger})/l$.  They form 
   a canonically conjugate pair,   $[\hat{\mathrm{y}}, \hat{\mathrm{x}} ] = \ii$.
   As $ \mathcal{P} r_x^2 \mathcal{P}^{\dagger}/l^2 - 
   \hat{\mathrm{x}}^2 = 1/2 \not =0$, the constant cyclotron motion energy term  
   $E_{\text{LL0}}=\hbar \omega_\mathrm{c}/2+\chi E_\mathrm{C} (1-\kappa^2)/2$ acquires a $\chi$-dependent correction due to electrostatic confinement by the isotropic part of the electric field gradient.
   The dimensionless ratio of characteristic energies $\chi =(U/E_\mathrm{C})^3$ is the parameter that controls the stability of the pairs.

 \subsection{Quantum state representation in the plane\label{sec:quantumstate}}

Projection onto the LLL maps the relative-motion state to the scaled complex plane of  $z=x+\ii y$. The same $(x,y)$ plane may be viewed both as the real-space  plane for the relative radius vector $\bm{r}$ and as the phase space of the canonically conjugate guiding-center coordinates. We will use this correspondence to characterize the  relevant states in two complementary ways: by the positions $z_n$ of their zeros, and by the Husimi $Q(z)$ function, as explained below.

The basis functions \eqref{eq:1eLaughlin} can be seen as projections of a coherent state onto the Fock basis, 
$\braket{ \alpha| m} = \sqrt{2 \pi} \varphi_m(z)$ where $\ket{\alpha}$ are  eigenstates of the annihilation operator $(\hat{\mathrm{x}} -\ii \hat{\mathrm{y}})/\sqrt{2} \ket{\alpha} =\alpha \ket{\alpha} $ with $\alpha = z^{\ast}/\sqrt{2}$.
According to  Eq.~\eqref{eq:psiLLL},
$\varphi(z) \equiv \braket{\alpha | \psi}/\sqrt{2 \pi}$ is identical to the wave function $\psi(\bm{r})$ projected to the LLL with length measured in units of $l$.
On the other hand, 
the absolute value squared of $\varphi(z)$ defines a positive phase-space quasiprobability density known as Husimi $Q$ function,
\begin{align} \label{eq:HusimiDef}
     Q(z) = \frac{1}{2 \pi} \lvert \braket{\alpha| \psi} \rvert^2 = |\varphi(z) |^2 \, .
   \end{align}
Husimi function is non-negative by construction and, as Eq.~\eqref{eq:HusimiDef} shows, admits direct physical interpretation as joint probability density for the commuting coordinates $r_x/l$ and $r_y/l$ for the physical wave function in the LLL.

The Wigner function $W(z)$ is the standard quantum counterpart of a classical
phase-space density \cite{Hillery1984DistributionFunctions} 
In
the same dimensionless variables as above, we define it in the
coordinate-like representation as
\begin{align}
 W(z) ={}& \frac{1}{\pi}
 \int_{-\infty}^{\infty} d\eta\,
 e^{\ii 2 x\eta}
 \langle\psi|y+\eta\rangle
 \langle y-\eta|\psi\rangle \, ,
 \label{eq:WignerDef}
\end{align}
where $\ket{y}$ is an eigenstate of $\hat{\mathrm{y}}$.  The Wigner function is widely used to represent spatiotemporal coherence in
electron quantum optics~\cite{Ferraro2013,Bisognin2019,Dubois2013,Fletcher2019}.
Its evolution is governed by the Moyal equation.  For a Hamiltonian that is at
most quadratic in the conjugate phase-space variables, the Moyal equation
reduces exactly to the classical Liouville equation; for a nonlinear
Hamiltonian the Liouville flow remains the leading semiclassical
approximation~\cite{Hillery1984DistributionFunctions}.  This correspondence underlies the phase-space description of
two-electron collisions and the modeling of partitioning statistics in
Refs.~\onlinecite{Pavlovska2023,Ubbelohde2023}.  The Husimi and
Wigner functions are members of the same Cahill--Glauber family of
quasiprobability distributions~\cite{CahillGlauber1969DensityOperators}.  
In these conventions, the Wigner and Husimi functions are related by
\begin{align}
   Q(z) & =
   \frac{1}{\pi} \int W(z') \, e^{-|z-z'|^2} d^2 z' \, .
   \label{eq:WignerQ}
\end{align}
Thus $Q$ is obtained by a minimum-uncertainty Gaussian smoothing of $W$.  It
retains the phase-space geometry of the state while removing Wigner
negativity, and in the present LLL problem it has the additional direct
probability-density interpretation established above.

The same LLL projection also gives the wave function a special analytic
structure~\cite{girvin1984formalism}.  Completeness of the Fock basis implies
that removing the universal factor $e^{-|z|^2/4}$ from $\varphi(z)$ leaves an
entire analytic function.  We refer to its zeros $z_n$ simply as the
wave-function zeros.  Since the Gaussian factor never vanishes, these are also
the zeros of $\varphi(z)$ and hence the points where $Q(z)=0$.  In modern
terminology, their constellation is the stellar (Segal--Bargmann)
representation of the quantum state~\cite{Chaubaud2020}.

For pure states these zeros provide an exact link to nonclassical phase-space
structure.  By Hudson's theorem, a pure state has an everywhere non-negative
Wigner function only if it is Gaussian~\cite{Hudson1974WignerNonnegative};
equivalently, a pure non-Gaussian state has zeros in its Husimi function
~\cite{Lutkenhaus1995NonclassicalSpace,Chaubaud2020}.  A zero of $Q$ therefore
implies that $W$ is negative somewhere, although the convolution in
Eq.~\eqref{eq:WignerQ} does not locate this negativity in the immediate
vicinity of the same $z_n$.  More importantly for the visualizations below,
the zero constellation encodes the non-Gaussian analytic factor nonlocally.
Together with the zero-free Gaussian envelope, which may include displacement
and squeezing, it specifies the state up to an irrelevant
overall phase.  Consequently, zeros in regions where the density itself is
very small may still be important for the shape of the wave function.  We
therefore plot both $Q(z)$ and the wave-function zeros $z_n$ to visualize the wave functions of quasibound states in the
LLL in Section~\ref{sec:resVis}.  In the full finite-field dynamics
of Section~\ref{sec:pairFormation}, the corresponding observables are the
physical density $|\psi(\bm{r},t)|^2$ and its isolated phase singularities, whose connection to $Q(z)$ and $z_n$ is approximate  rather than exact due to residual higher-Landau-level admixture $\psi^{\perp}(\bm{r}, t)$.

\section{Stability diagram and the pair decay rates\label{sec:stabilityAndRates}}
Internal state of a Laughlin pair is described by the Hamiltonian \eqref{eq:projected}. In a flat potential ($\chi=0$)
the angular momentum quantum number $m$ is conserved and the pairs of all radii are possible. For finite $\chi$, the 
maximal index $\bar{m}_{\text{max}}$ of a quasibound state can be estimated using a semiclassical argument as follows.
The saddle points of the effective potential for the relative motion, $\bm{r}=(0, \pm d_0)$, correspond to $\pm \mathrm{y}=d_0/l=\chi^{-1/3}/\sqrt{2}$ and energy $\varepsilon_{\text{th}} = (3/2) \chi^{1/3}$ in dimensionless units.  
The phase-space area available to closed classical trajectories is finite
 for $\chi>0$. In our units where
$[\hat{\mathrm y},\hat{\mathrm x}]=\ii$, dividing this 
area by $(2\pi)$, which corresponds to one Planck constant,  gives the semiclassical number of states.
For the Coulomb potential this calculation gives
\begin{align} \label{eq:mmax}
 \bar{m}_{\text{max}} = \mathcal{A}_{\kappa} \chi^{-2/3} \, ,
\end{align}
where the prefactor $\mathcal{A}_{\kappa}$ depends only on the geometry parameter $\kappa$ via $u_0 \equiv \kappa^{-1} \sinh [(\sinh^{-1} \kappa )/3]$,
\begin{multline}
    \mathcal{A}_{\kappa} = 
    \frac{3}{\pi} \int_{u_0}^{1/2}\frac{u \,du}{\sqrt{(u+1)\,\big(4\kappa^2 u^3+3 u-1\big)}} \\
     =\frac{1}{2 \pi} \begin{cases}
        (3 - 2 \ln 2)/\sqrt{3} =0.93167 \, ,&  \kappa=0  ,\\
        0.8049061497 \, , & \kappa=1, \\
       3 \, \ln (2 +\sqrt{3}) / \kappa= 3.95087 /
       \kappa \, , & \kappa \to \infty .
    \end{cases}
\end{multline}

The energies $\varepsilon_m^{(0)}$ of the unperturbed ($\chi =0$) Laughlin pairs   are continuously connected at $\chi >0$ to the poles of the resolvent $(\varepsilon- \mathrm{H}+ \ii 0)^{-1}$ (the complex resonances) at $\varepsilon=\varepsilon_m - \ii \gamma_m$. The energies $\varepsilon_m$, decay rates $2 \gamma_m$ (in units of $E_\mathrm{C}/\hbar=\sqrt{\omega_\mathrm{c} \omega^{\star}})$ and the wave functions of the corresponding eigenstates $\mathrm{H} \ket{\varepsilon_m} =(\varepsilon_m - \ii \gamma_m) \ket{\varepsilon_m}$ are  computed numerically using the complex-scaling method described in Appendix \ref{app:Scaling}.

\begin{figure}
    \includegraphics[width=0.499\textwidth]{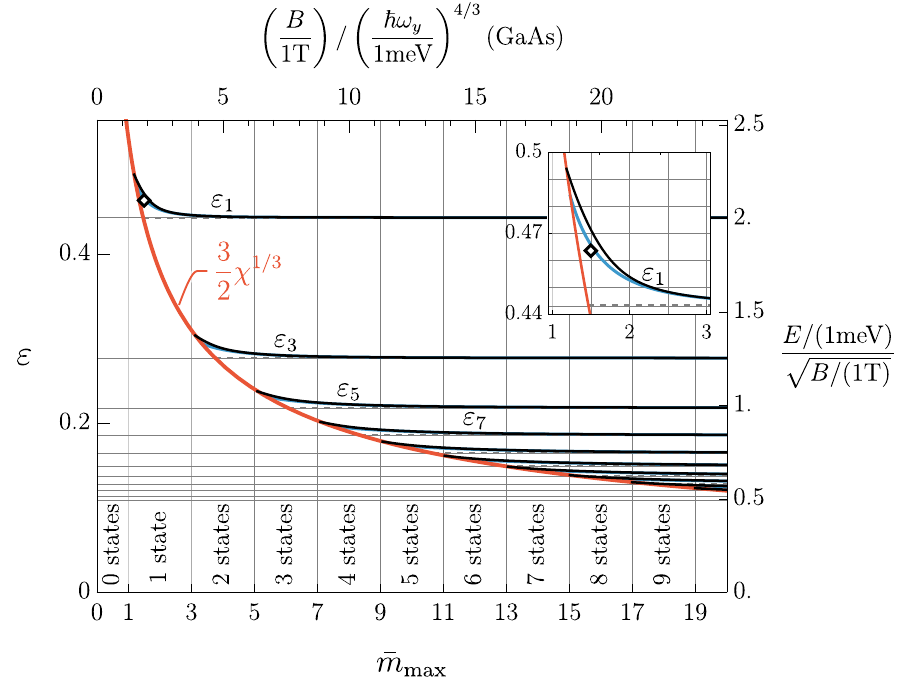}
 \caption{Dimensionless energies $\varepsilon_m=E_m/E_\text{C}$ of quasibound Laughlin pairs in a purely quadrupole  ($\kappa=1$) electric field gradient in the LLL approximation. No resonances are present below the threshold energy $\varepsilon_\mathrm{th}=(3/2)\chi^{1/3}$ (orange line).   
    Numerical results obtained by the complex-scaling method (black line) agree well with the second-order perturbation theory (blue line), Eq.~\eqref{eq:eps2ndorder}.
    Horizontal lines mark the unperturbed energies $\varepsilon_m^{(0)}$ of pairs in the bulk.
     The inset shows a magnified view of the $m=1$ state near the threshold and
    the diamond indicates the energy of the pair state observed in the time-dependent wave-packet calculation (see Section~\ref{sec:pairFormation} below).
    } 
    \label{fig:energiesCS1Dk1}
\end{figure}

Figures \ref{fig:energiesCS1Dk1} and \ref{fig:energiescomplexpartCS1Dk1} illustrate the main results of this section.
We find that the condition
\begin{align} \label{eq:threshold}
    \varepsilon_m > \varepsilon_{\rm{th}} \, = (3/2) \chi^{1/3} \, ,
\end{align}
shown by the orange line in Figure~\ref{fig:energiesCS1Dk1},
accurately describes the threshold values of the main parameter $\chi$ for the existence of the pair states and the number of quasibound states is well approximated by  $\lfloor  \bar{m}_{\text{max}} \rfloor$.  (As we focus on the spin-triplet states, only odd $m$ are included in the figure.) 
Above the threshold \eqref{eq:threshold}, the energies quickly converge to the bulk limit,  $\varepsilon_m \to \varepsilon_m^{(0)} $ as $\chi \to 0$ (as indicated by  horizontal lines in Figure~\ref{fig:energiesCS1Dk1}), and can be accurately estimated by perturbation theory. 
The isotropic and the quadrupole part of the electric field gradient contribute to the first and the second order  in $\chi$, respectively, as follows~\footnote{Note that in our notation the isotropic part $\propto (1-\kappa^2)$ also contributes to a constant shift in $E_{\text{LL}0}$.}:
\begin{equation} \label{eq:eps2ndorder}
     \varepsilon_m=
     \varepsilon_m^{(0)}+\chi \frac{1-\kappa^2}{2}(2m+1)+ \chi^2 \frac{\left(1+\kappa^2 \right)^2}{4}
     c_m +O(\chi^3) \, ,
\end{equation}
where
\begin{equation}
    c_m=\frac{(m+1)(m+2)}{\varepsilon^{(0)}_m-\varepsilon^{(0)}_{m+2}}  
  +\frac{m(m-1)}{\varepsilon^{(0)}_m-\varepsilon^{(0)}_{m-2}}  \, .
\end{equation}

As the comparison to numerically exact results in Figure \ref{fig:energiesCS1Dk1} shows, Eq.~\eqref{eq:eps2ndorder} gives an accurate approximation of the quasibound state energy  over the entire range of its existence.
The upward shift of $\varepsilon_m$ near the threshold seen in  Figure \ref{fig:energiesCS1Dk1} is the energy of the induced quadrupole moment of the pair in the external field gradient (the quadratic term of the expansion \eqref{eq:eps2ndorder}).

\begin{figure}
 \includegraphics[width=0.48\textwidth]{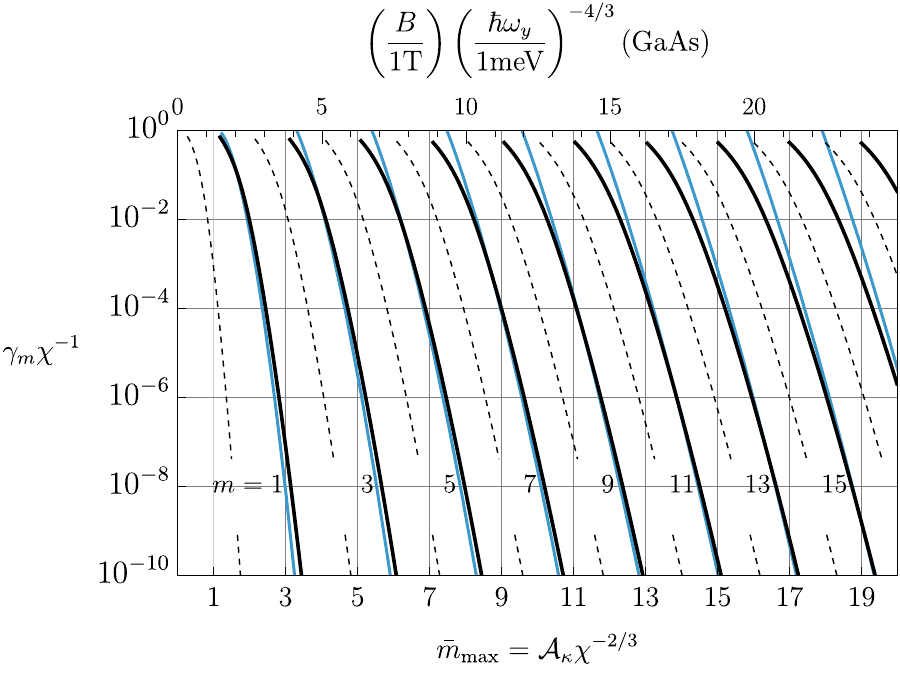}
\caption{
Dimensionless resonance halfwidths, scaled as
$\gamma_m/\chi$, for pure-quadrupole anisotropy  $\kappa=1$, plotted as
functions of $\mathcal{A} _1\chi^{-2/3}$ on the lower axis. Solid black curves are
 numerical complex-scaling results for the odd-$m$ branches indicated by
 the labels; dashed black curves show the even-$m$ branches. Blue curves
 show the closed-form tunneling estimate in Eq.~\eqref{eq:gammam} for the
 odd-\(m\) branches. The upper axis gives the corresponding  parameter
 combination for GaAs devices.
\label{fig:energiescomplexpartCS1Dk1}}
\end{figure}

The decay rates as a function of $\bar{m}_{\text{max}}$, shown in Figure~\ref{fig:energiescomplexpartCS1Dk1} for $\kappa=1$,   start at $\gamma_m \sim \chi$ at the threshold as the $m$-th resonant state becomes well-defined at $\bar{m}_{\text{max}}  \approx m$. As the distance $\varepsilon_m-\varepsilon_{\text{th}}$ from the threshold is increased with growing $\bar{m}_{\text{max}}$, the decay rates $\gamma_m$ are quickly reduced.  This stabilization is captured analytically by
\begin{align}
    \gamma_m^{\text{tun}}=
    \frac{1}{ \pi} \left ( 
    \varepsilon_m^{(0)} \right )^3  \exp \left [ -  \frac{\pi \left ( \varepsilon_m - \varepsilon_{\text{th}}  \right ) }{ \chi \sqrt{3 \,(1+\kappa^2)} }   \right ] \, ,
    \label{eq:gammam}
\end{align}
where the attempt frequency is  estimated from the rotation period of an unperturbed pair, the decay probability is approximated by the transmission probability under  a parabolic potential near the saddle points~\cite{Pavlovska2023,Silvestrov2025}, and the energies $\varepsilon_m$ are computed from Eq.~\eqref{eq:eps2ndorder}.   (Note that this 
parabolic-saddle estimate is not asymptotically exact as $\chi\to0$ for a fixed $m$.)

Figures~\ref{fig:energiesCS1Dk1} and
\ref{fig:energiescomplexpartCS1Dk1} illustrate the $\kappa=1$ case.
For general $\kappa$, the dependence on geometry is retained explicitly
through the prefactor $\mathcal{A}_{\kappa}$ in
$\bar{m}_{\text{max}}=\mathcal{A}_{\kappa}\chi^{-2/3}$, as well as through
the remaining $\kappa$-dependent factors in
Eqs.~\eqref{eq:eps2ndorder} and~\eqref{eq:gammam}.
The special $\kappa=0$ case is illustrated in the Appendix~\ref{app:kappazero}.

The exponential in Eq.~\eqref{eq:gammam} gives an estimate for the decay rate ratio between the neighboring states  $\ln(\gamma_m/\gamma_{m-1}) \simeq C_\kappa (m/\bar m_{\text{max}})^{-3/2}$, with
$C_\kappa=\pi/[4\mathcal A_\kappa^{3/2}\sqrt{3(1+\kappa^2)}]$. These relatively large values (e.g., $C_{\kappa=1} \approx  7.0$), as well as non-perturbative numerical results shown in Figure~\ref{fig:energiescomplexpartCS1Dk1}, confirm the large separation in tunnel coupling strengths of the Laughlin pairs of alternating symmetry at a saddle point potential. 
When both spin sectors are available, this strong separation of even- and
odd-parity decay rates provides the lifetime contrast required for the
time-domain singlet--triplet filtering scheme proposed in
Ref.~\cite{Silvestrov2025}.

\section{Visualization of resonance wave
    functions \label{sec:resVis}}

In Section~\ref{sec:stabilityAndRates} we characterized the  resonances corresponding to unstable Laughlin pairs by their
complex energies.  We now examine their wave functions in order to see how an
ideal Laughlin pair is deformed by the guiding potential and how a quasibound
state connects to the outgoing scattering channels.  At $\chi=0$ the states
$\varphi_m(z)$ in Eq.~\eqref{eq:1eLaughlin} have angular momentum $m\hbar$.
For $\chi>0$ the quadrupole term mixes different angular momenta, so $m$ is no
longer a conserved angular-momentum quantum number.  We continue to use it as
a robust branch label: the resonance denoted by $m$ is obtained by continuous
evolution from the corresponding rotationally symmetric Laughlin state as
$\chi$ is increased.

We visualize the resonant eigenstates  computed with the complex-scaling
method of Appendix~\ref{app:Scaling} by their Husimi density $Q(z)$ [Eq.~\eqref{eq:HusimiDef} with $\ket{\psi}=\ket{\varepsilon_m}$] and by its
zeros $z_n$.  These provide two complementary views of the same wave function.  As
discussed in Section~\ref{sec:quantumstate}, $Q(z)$ is directly the
relative-coordinate probability density in the LLL: it tells us where one
electron is likely to be found with respect to the other.  The same plane is
also the phase space of the guiding-center motion.  In this interpretation a
well-confined stationary state is expected semiclassically to resemble a
quantum-broadened, phase-averaged distribution around a closed classical
orbit.  Its density accumulates where the classical motion slows down, while
for a resonance it can also reveal the outgoing branches along which the pair
decays.

The zeros expose a different and particularly stable part of the analytic
structure.  They move continuously as the parameters of a resonance are
varied and connect the numerical states to two analytically transparent
limits.  For the pure Coulomb problem,
$\varphi_m(z)\propto z^m e^{-|z|^2/4}$, so its Gaussian-stripped analytic part
has a zero of order $m$ at the origin.
In the opposite, Coulomb-free saddle problem, the exact scattering
eigenfunctions are parabolic-cylinder functions whose complex zeros organize
along anti-Stokes lines~\cite{complexzerossegura2013}.  These two limits
provide reference patterns for interpreting the numerical zeros below.

Figure~\ref{fig:WFresonances} applies these complementary views to four
resonances in a traceless electric field gradient (i.e., $\kappa=1$ symmetric saddle).  The top row follows the $m=1$ branch and
the bottom row the $m=3$ branch.  Within each row, the left panel is close to
the dissociation threshold, while the right panel shows a better-confined
state of the same branch.  This arrangement lets us first identify the
deformation of a well-confined Laughlin pair and then see what changes as the
resonance approaches the threshold.
For direct comparison with the phase-space representation commonly used in
 electronic-coherence tomography~\cite{Bisognin2019,Fletcher2019}, Appendix~\ref{sec:WignerFunctions} also shows the
  Wigner quasiprobability distributions for the same four resonances. Their sign-alternating
  interference fringes expose the fine coherence structure suppresed by the Gaussian
 smoothing in Eq.~\eqref{eq:WignerQ}, whereas the Husimi density shows a clearer picture of  the
   large-scale closed-orbit and escape-channel geometry and the zeros compactly
   encode the non-Gaussian analytic structure.
\begin{figure*}[htbp]
  \setlength{\tabcolsep}{2pt}
  \begin{tabular}{cc}
      \begin{overpic}[width=0.48\textwidth]{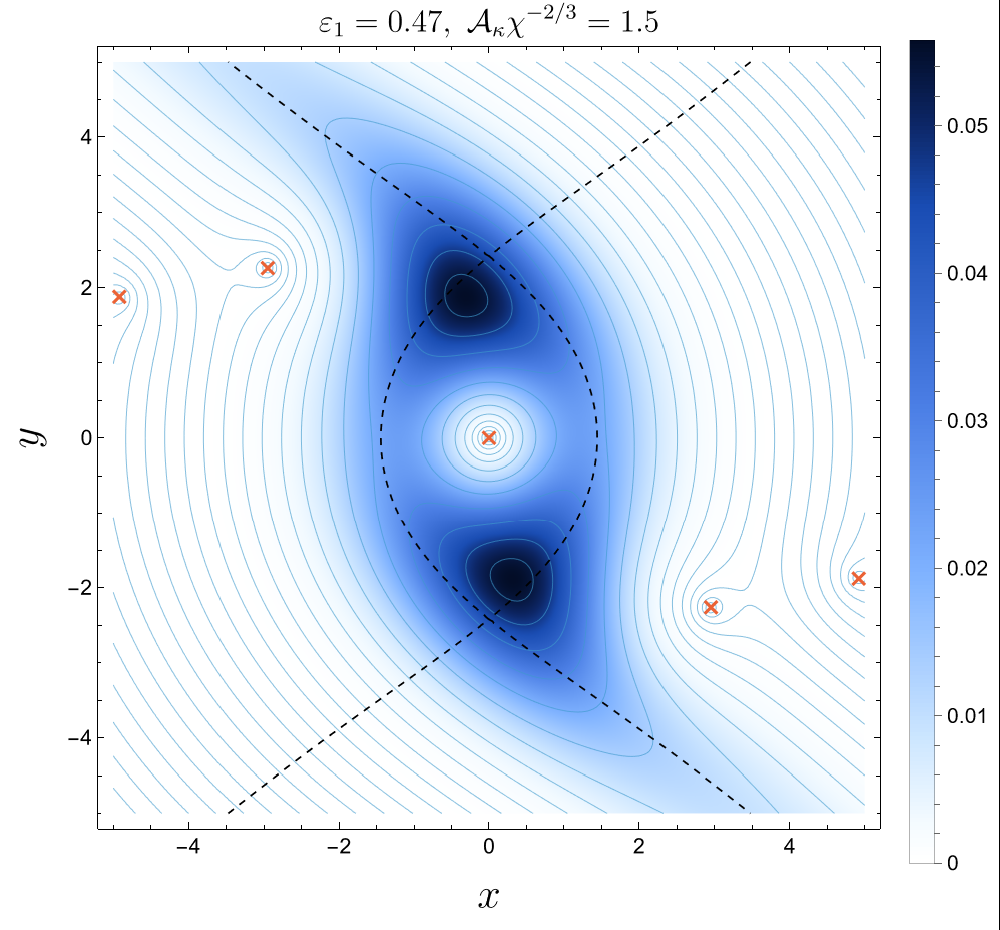}\put(2,90){\footnotesize\textbf{(a)}} \end{overpic} &
      \begin{overpic}[width=0.48\textwidth]{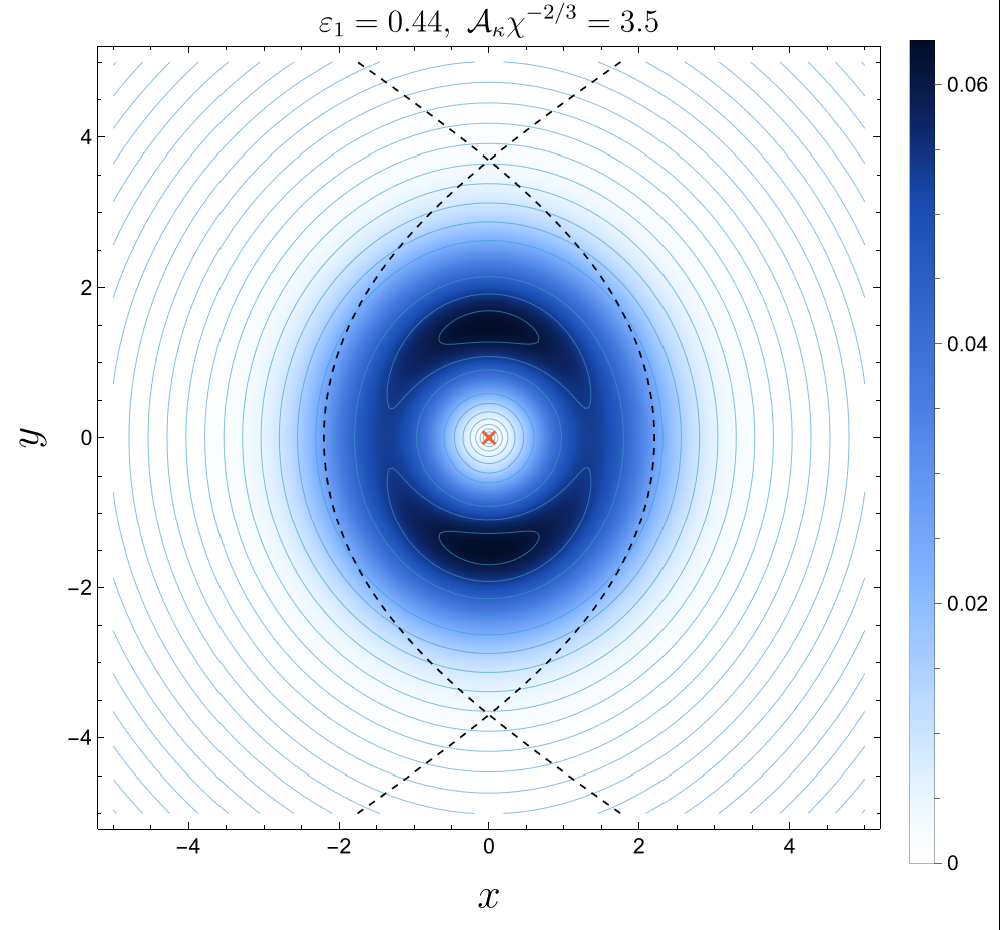}\put(2,90){\footnotesize\textbf{(b)}}\end{overpic} \\
      \begin{overpic}[width=0.48\textwidth]{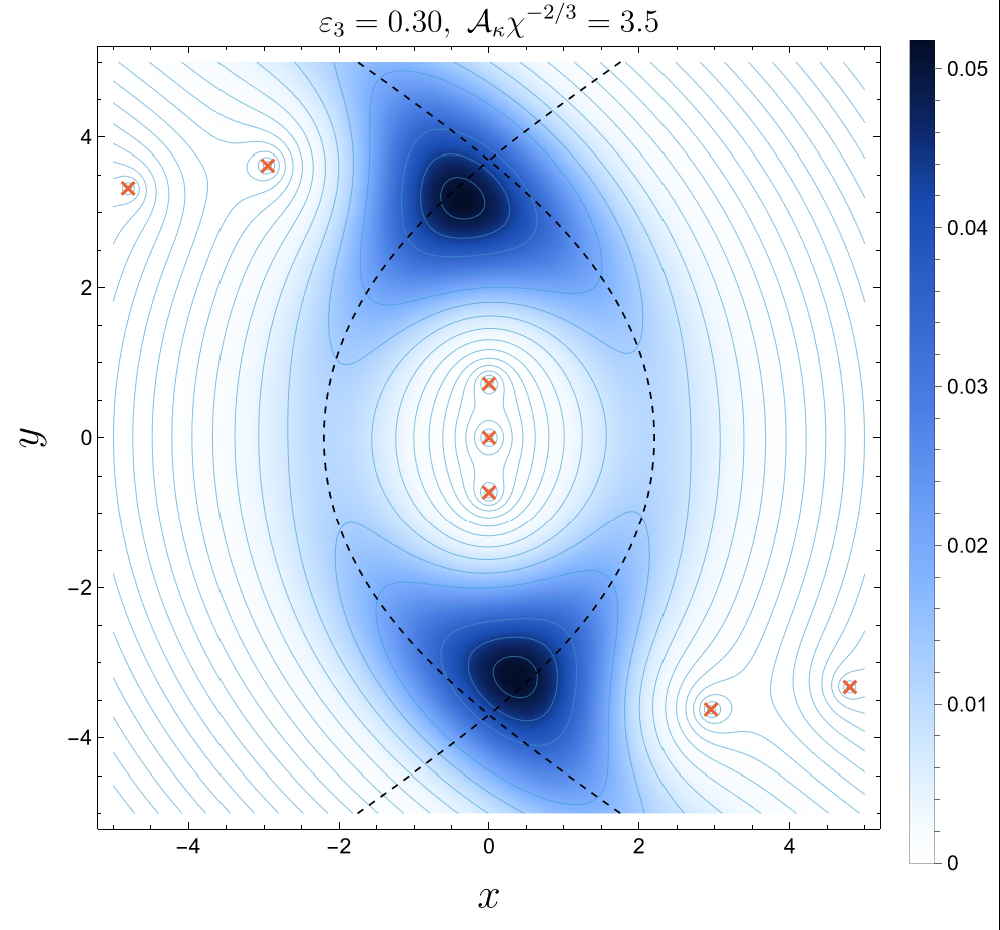}\put(2,90){\footnotesize\textbf{(c)}}\end{overpic} &
      \begin{overpic}[width=0.48\textwidth]{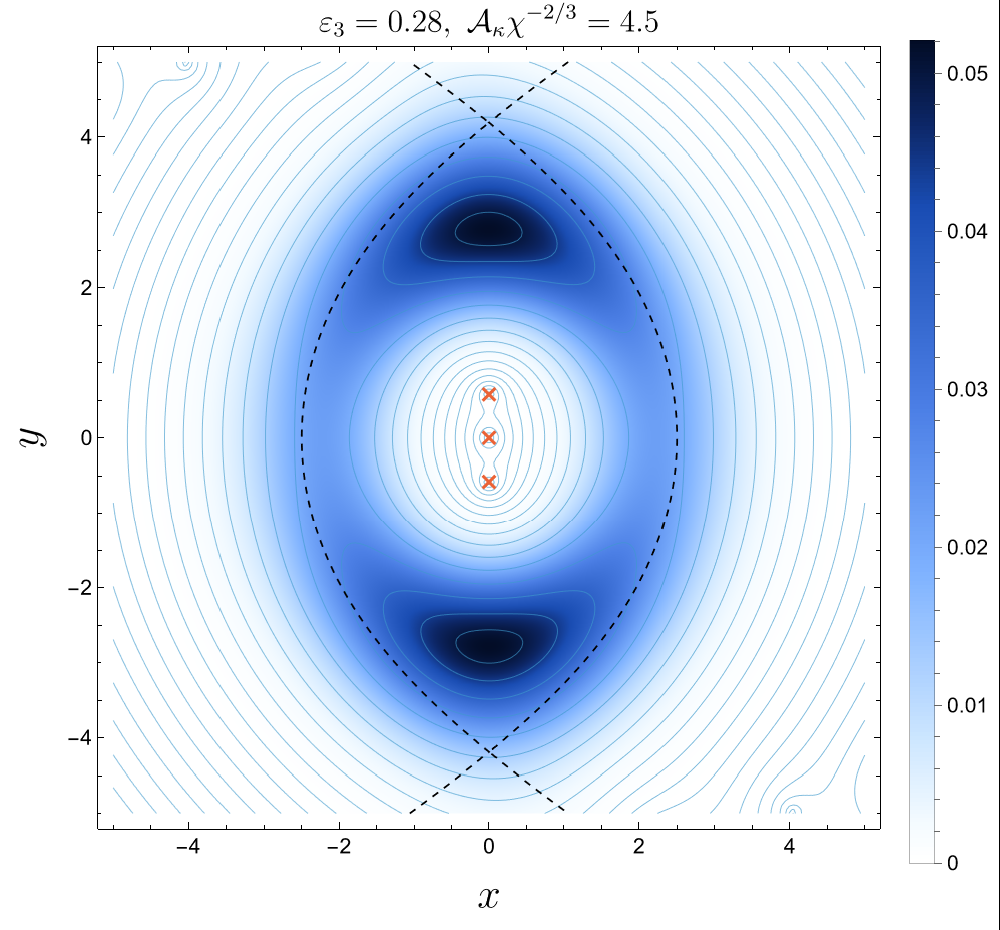}\put(2,90){\footnotesize\textbf{(d)}}\end{overpic}
  \end{tabular}
  \caption{Husimi density of Laughlin-pair resonances for $\kappa=1$.  Panels~(a) and (b) show the $m=1$
  branch at $\bar m_{\mathrm{max}}=1.5$ and $3.5$, respectively; panels~(c)
  and (d) show the $m=3$ branch at $\bar m_{\mathrm{max}}=3.5$ and $4.5$.
  The corresponding resonance energies are printed above the panels.  Blue
  shading shows $Q(z)$, and the solid contour levels are chosen to be equally
  spaced for a Gaussian profile.  Crosses mark its zeros.  The dashed curves
  are the classical critical level at
  $\varepsilon=\varepsilon_{\mathrm{th}}$, separating the closed central
  trajectories from the incoming and outgoing branches.  The better-confined
  states in panels~(b) and (d) retain an approximately bulk-like distribution,
  modified by density pile-up near the interaction-induced saddle points and
  by splitting of the higher-order zero.  The near-threshold states in
  panels~(a) and (c) additionally show outgoing density tails and outer zeros
  associated with the scattering structure.\label{fig:WFresonances}}
\end{figure*}

The better-confined resonances in the right-hand panels provide the simpler
starting point.  Their energies lie well above the
dissociation threshold and their Husimi densities remain concentrated inside
the classical critical level.  Panels~(b) and (d) consequently retain much of
the higher symmetry and annular form of the bulk Laughlin states.  The main
departure visible in the density is the pile-up above and below the origin,
near the interaction-induced saddle points $\bm r=(0,\pm d_0)$.  

The zero pattern reveals a second departure from the bulk state.  For the
$m=1$ branch the simple zero remains at the origin [panel~(b)].  For $m=3$,
the quadrupole perturbation lifts the degeneracy of the third-order zero: one
zero remains pinned to the origin by odd exchange symmetry, while the other
two move symmetrically away from it [panel~(d)].  More generally, inversion
symmetry requires nonzero zeros to occur in $\pm z_n$ pairs.  Thus the $m$
zeros inherited from the order-$m$ zero of the ideal Laughlin state remain
recognizable within the quasibound structure even though $m$ itself no longer
denotes angular momentum.

The near-threshold resonances in the left-hand panels combine this closed-orbit
structure with that of a stationary scattering state.  For the $m=1$ state at
$\bar m_{\mathrm{max}}=1.5$ and the $m=3$ state at
$\bar m_{\mathrm{max}}=3.5$, the energies $\varepsilon_1=0.471$ and
$\varepsilon_3=0.295$ lie only just above the respective threshold values
$0.438$ and $0.287$. 
The Husimi density then loses the approximate bulk
symmetry: it becomes concentrated near the two interaction-induced saddle
points and develops pronounced outgoing tails along the anti-diagonal
asymptotes of the critical level.  In phase-space language, the same
stationary distribution contains both the quantum-broadened remnant of the
closed trajectories and the escape flow of the scattering ensemble. 
Only inversion symmetry remains, reflecting particle indistinguishability.

The zeros display the same coexistence in a sharper form.  The single central
zero in panel~(a), and the three central zeros in panel~(c), are the continuous
descendants of the corresponding bulk Laughlin zero.  Additional zeros appear
in the low-density outer region.  They form inversion-related pairs and, away
from the Coulomb-dominated center, organize towards the anti-Stokes lines of
the saddle scattering solutions.  These outer zeros describe the analytic
structure of the outgoing wave and do not change the identity of the
resonance branch.

Panel~(a) provides the stationary reference used in the dynamical calculation
of Section~\ref{sec:pairFormation}.  At $\bar m_{\mathrm{max}}=1.5$ the only
supported odd branch is $m=1$; its near-threshold resonance has two density
maxima near the interaction-induced saddle points and one central,
exchange-enforced zero.  The density records where the quasibound pair resides
and how it escapes, while the internal zero preserves its continuity with the
ideal Laughlin pair.  Together these features give the structural comparison
used below to identify the late-time remnant of a two-electron collision.

\section{Pair formation in dynamical two-electron collisions\label{sec:pairFormation}}

The resonances discussed in Sections~\ref{sec:stabilityAndRates} and
\ref{sec:resVis} are stationary states of the relative motion.  We now ask
whether such a state can be populated dynamically in a collision of two
spin-polarized electrons at a saddle-point potential, and how its formation is
reflected in the time evolution of the wave function.  We solve the
time-dependent Schrödinger equation (TDSE) for the full finite-field
relative-coordinate wave function $\psi(\bm r,t)$ governed by
$\mathcal H_{\bm r}$, Eq.~\eqref{eq:Hr}, without projection to the lowest
Landau level.  
This complements the stationary LLL calculation and develops
the tunneling-formation scenario of Ref.~\onlinecite{Silvestrov2025} beyond
the auxiliary one-dimensional double-barrier model used there: here both the
Coulomb interaction and the two-dimensional saddle dynamics are retained in
the propagated Hamiltonian.

The two electrons are initialized far apart on opposite incoming branches of
the saddle, where their Coulomb interaction is negligible and they may be
interpreted as independently emitted wave packets.  We choose two 
single-electron Gaussians with the same full shape (compatible with injection into the LLL)
so that their product wave function
factorizes exactly into center-of-mass and relative-coordinate Gaussians. 
For the symmetric collision considered here the center-of-mass packet is centered
at $\bm R=0$ and is decoupled from the Coulomb interaction; only the
relative-coordinate factor $\psi(\bm r,t)$ needs to be propagated numerically.
The orbital state is antisymmetrized to describe spin-polarized electrons.  Its two
Gaussian branches are initially centered close to the incoming separatrix and
far enough from the interaction region so that the Coulomb energy is negligible.
The construction of this factorizable state in the exact Coulomb-free saddle
basis is described in Appendix~\ref{app:TDSEintial}.

We consider a symmetric saddle, $\kappa=1$, with $\chi=0.025$ and
$\omega_{\mathrm c}/\omega^\star=25/4$.  These parameters give
$\bar m_{\mathrm{max}}=1.5$, so that the spin-polarized stationary problem
supports only the odd $m=1$ resonance.

 The incoming state has mean dimensionless energy
$\bar\varepsilon=0.441$, close to the dissociation threshold
$\varepsilon_{\rm th}=0.438$; its finite energy width overlaps the
$m=1$ resonance. In the finite-field time-dependent calculation, we
extract the resonance energy $\varepsilon_1=0.463$ from the Fourier
spectrum of the overlap between the evolving state and the late-time
 remnant~\cite{FeitFleck1982}. This value, marked by the diamond in the
 inset of Fig.~\ref{fig:energiesCS1Dk1}, lies below the LLL complex-scaling
 result $\varepsilon_1=0.471$, consistent with the higher-Landau-level
 admixture retained in the full calculation.

The relative-coordinate state is
propagated on a two-dimensional grid by a split-step Fourier method; its
initial parameters and the numerical implementation are specified in
Appendix~\ref{app:TDSE}.  The complete evolution is shown in the Supplemental
Material~\cite{video:tdsepropagation}, and representative snapshots are shown
in Fig.~\ref{fig:Fig_Hussimi8snapshots}.

The density and its zeros carry complementary information throughout the
collision.  The blue map in Fig.~\ref{fig:Fig_Hussimi8snapshots} shows the
full finite-field probability density $|\psi(\bm r,t)|^2$, while the crosses
mark isolated density zeros, equivalently phase singularities of
$\psi(\bm r,t)$.  In the strong-field regime their positions can be compared
with the zeros of the LLL-projected Husimi function discussed in
Section~\ref{sec:quantumstate}, although this correspondence is no longer an
exact identity for the unprojected wave function. 
Because of the statistics-imposed odd exchange symmetry, 
the initial state of the relative coordinate is a Schrödinger-cat state,
i.e., a coherent superposition of two distant Gaussians.
 The coherent nature of this superposition is manifested by a dense straight line of equally spaced Husimi zeros.
 (The corresponding incoherent mixture of the same two Gaussian branches has  a strictly positive Husimi
function with no zeros.)  The positions of these exchange-symmetry-dictated zeros 
are computed using the analytic  LLL approximation \eqref{eq:zerosLLL} (see Appendix \ref{app:TDSEintial}) and are shown in
 panel~(a). The interzero spacing  is inversely
 proportional to the initial distance before the collision.

\begin{figure*}
    \begin{overpic}[width=0.99\textwidth]{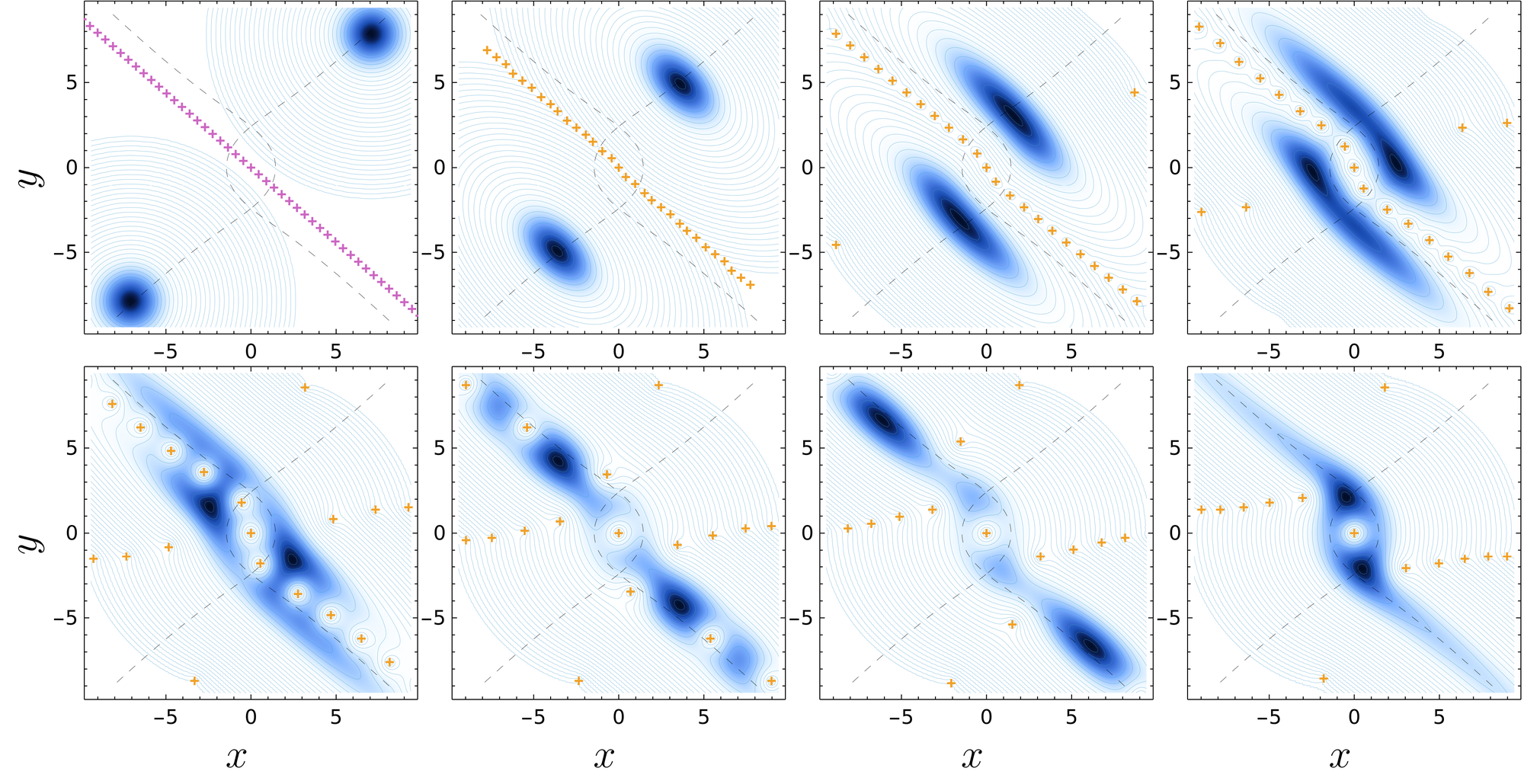}
     \put(3.5,52){\footnotesize\textbf{(a)}}
     \put(27.5,52){\footnotesize\textbf{(b)}}
     \put(51.5,52){\footnotesize\textbf{(c)}}
     \put(75.5,52){\footnotesize\textbf{(d)}}
     \put(3.5,28){\footnotesize\textbf{(e)}}
     \put(27.5,28){\footnotesize\textbf{(f)}}
     \put(51.5,28){\footnotesize\textbf{(g)}}
     \put(75.5,28){\footnotesize\textbf{(h)}}
    \end{overpic}
    \caption{Formation of a Laughlin-pair resonance in a collision of two
    electrons.  The panels show the relative-coordinate probability density $|\psi(\bm r,t)|^2$ (blue shading
    and contours) at dimensionless times
    $\tau=t \, E_{\mathrm C}/\hbar=0,11,24,32,40,53,62,$ and $242$ in panels
    (a)--(h), respectively.  Each panel is normalized independently to its own density maximum.
     Crosses mark isolated zeros of the wave
    function.  In panel~(a) their straight-line arrangement is calculated
    analytically in the LLL limit from the antisymmetrized initial Gaussian
    state; in the later panels the zeros are obtained numerically.  The dashed
    curves are the exact critical level lines at
    $\varepsilon=\varepsilon_{\text{th}}$.  In the central region they go
    around the Coulomb singularity and connect the interaction-induced saddle
    points; far from the center, their incoming and outgoing branches approach
    the main and anti-diagonal, respectively.
    Panels~(a) and (b) show the incoming cat state and its predominantly
    quadratic squeezing--stretching dynamics.  Panels~(c)--(f) show the
    interaction-induced deformation, passage between the two
    interaction-induced saddle points, and separation of the prompt
    scattering components.  In panels~(g) and (h) these components leave the
    central region and expose a persistent odd remnant with one central, exchange-enforced zero.  In
    the quasibound region, its final density profile agrees qualitatively with
    the stationary $m=1$, $\bar m_{\mathrm{max}}=1.5$ resonance in
    Fig.~\ref{fig:WFresonances}(a).}
    \label{fig:Fig_Hussimi8snapshots}
\end{figure*}
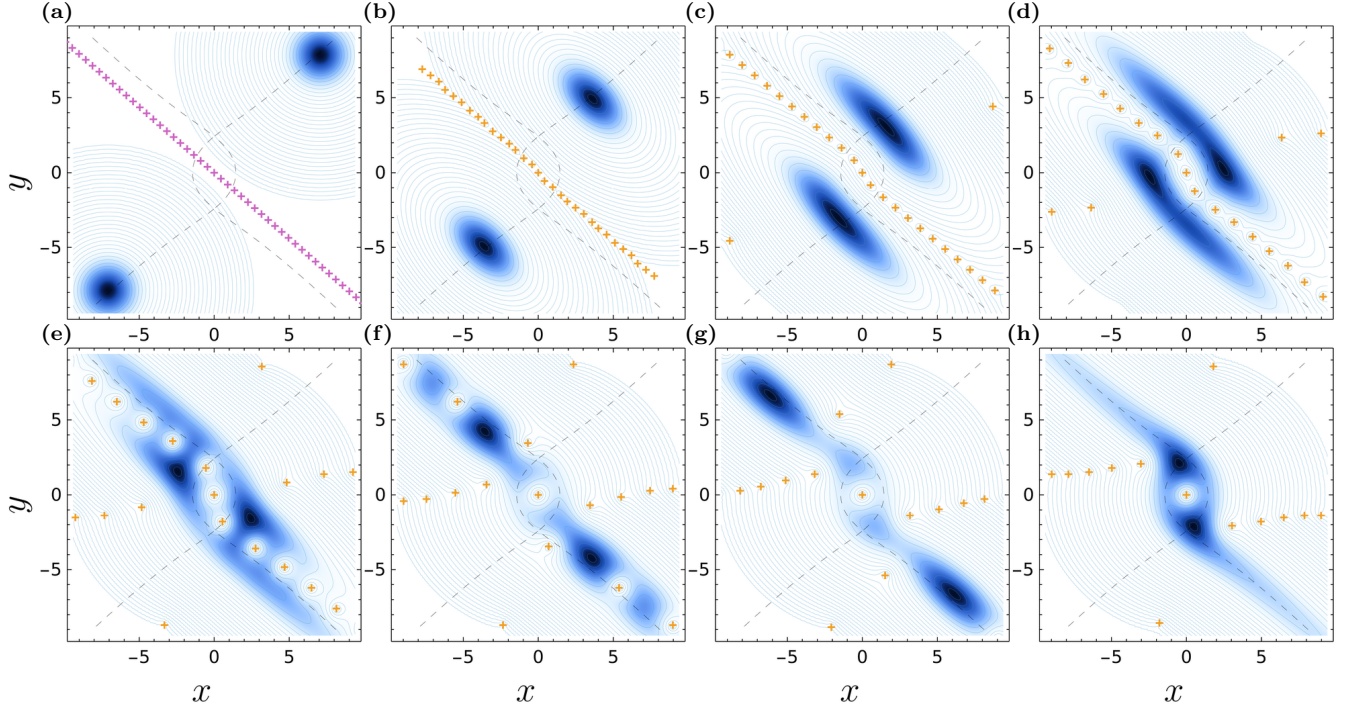

The evolution in Fig.~\ref{fig:Fig_Hussimi8snapshots} can be separated into
three stages.  Initially, the two well-separated density maxima and the
straight line of antisymmetrization zeros form the expected odd cat
state [panel~(a)].
During the early approach [panel~(b)] the density is
predominantly squeezed and stretched by the Liouvillian flow of the quadratic
saddle Hamiltonian, before the Coulomb interaction substantially changes its
shape.  This part of the evolution retains the two-Gaussian character and
provides the simple reference motion against which the later non-Gaussian
features can be identified. 

The first such features appear in panel~(c), where the density deforms in the
Coulomb-dominated region and an additional pair of zeros
enters the plotted area from large distance.  These zeros do not belong to the
initial antisymmetrization chain, but are expected to accommodate deviations from Gaussianity. 
In panel~(d) the deformation becomes
pronounced: the principal density maximum flows from the first
interaction-induced saddle point towards the second, while the part reflected
at the first saddle begins to leave the interaction region.  To leading
semiclassical order, this separation into two transient components can still
be understood as Liouvillian flow of the guiding-center Wigner density.  In
the strong-field limit the plotted spatial density approaches the
corresponding Husimi distribution, which adds the Gaussian smoothing described
in Section~\ref{sec:quantumstate}.
The later snapshot in panel~(e) shows the limits of this classical
phase-space dynamics picture.  The density does reflect 
the stretched, accelerating fraction along the anti-diagonal in the upper and the lower quadrants and a
pronounced pile-up of the other transient component near the interaction-induced saddles, on the opposite side of the classical separatrix.   However, the squeezing of the density envelope towards the escape
separatrix has saturated, and the non-classical zeros now lie within the density
envelope and enforce substantial modulation beneath it.
The zeros in this intermediate regime cannot
be assigned uniquely to the incoming cat state or to the final resonance.  By
panel~(f), two outgoing density maxima are resolved and still present in the frame, with a
single zero separating them on each of the two inversion-related outgoing
branches. We interpret these components as classical subensembles corresponding to  asymptotic   $\exp(-\lambda_I t)$ and
$\exp(-\lambda_I t/2)$ escape laws derived in Ref.~\onlinecite{Silvestrov2025} for phase-space trajectories approaching the saddle point once or twice, respectively. (Here 
\(\lambda_I\) is the Lyapunov exponent of either interaction-induced saddle.)

A qualitatively different separation becomes visible in panels~(g) and (h).
In panel~(g), one large-weight classical component continues to escape, while
a much smaller central component with one isolated zero remains in the
interaction region.  Once the prompt components have left, the normalized
central pattern ceases to change appreciably and only this remnant is visible
in the selected region at $\tau=242$ [panel~(h)].  Its two density maxima lie
near the interaction-induced saddle points, and its single zero in the
quasibound core is the one required by the odd exchange symmetry.  This
qualitatively reproduces the central structure of the independently
calculated $m=1$ resonant state
in Fig.~\ref{fig:WFresonances}(a), evaluated at the same
$\bar m_{\mathrm{max}}=1.5$.  The comparison is restricted to the quasibound
core: the finite wave packet also contains outgoing components and is absorbed
at the remote grid boundary, whereas the stationary resonance satisfies
outgoing boundary conditions.

The characteristic time for the classical dynamics near either
interaction-induced saddle is the inverse of  $\lambda_I$ ($\omega^{(2)}$ in the notation of \cite{Pavlovska2023}).  In the present
units this time is
\begin{align}
    \tau_I=\frac{\chi^{-1}}{2\sqrt{3(1+\kappa^2)}}=8.2 \, ,
    \label{eq:interactionTimeSectionV}
\end{align}
which approximately corresponds to the chosen time spacings between snapshots in panels (a) up to (g) in Fig.~\ref{fig:Fig_Hussimi8snapshots}.

The stationary halfwidth $\gamma_1$ in
Fig.~\ref{fig:energiescomplexpartCS1Dk1} gives a probability-decay time
$\tau_1=(2\gamma_1)^{-1}\approx88\approx11\,\tau_I$, where we used
$\gamma_1/\chi=0.227$ from the stationary calculation.  This separation of
time scales allows the prompt classical fractions to leave before the resonant
component decays. 

In the numerical TDSE calculation, exponential decay dominates the tail of the total probability $P(\tau)$ remaining on
the numerical grid (i.e.,
remaining after application of the smooth absorbing
    mask
with the radius of $r_{\text{mask}} \approx 40$ 
as described in Appendix \ref{app:TDSEpropagation}).
Fitting $P(\tau) \propto \exp( - \tau/\tau_1)$ for
$387 > \tau > 144$ gives 
$\tau_1= 56.5$, a somewhat shorter lifetime than  the  stationary LLL estimate. This discrepancy is consistent with energy correction  due to higher Landau levels: using the unrounded energies,
the downward shift of  $\Delta \varepsilon_1 =7.87 \cdot10^{-3}$  
in the exponential energy dependence in Eq.~\eqref{eq:gammam} gives a factor of $1.50$ enhancement to  $\gamma_1$. This  brings the LLL value of $\tau_1=88$ down to $58.4$, close to the time-dependence fit.

 The probability  $P(\tau)$ is a proxy for the probability of the electrons to stay close, and thus can be used to roughly estimate the yield of Laughlin pair creation.
Extrapolating the exponential tail of $P(\tau)$  backwards and taking a
$1\%$--$10\%$ relative
deviation from it as the onset of the exponential regime, gives the onset time
$115\leq\tau_0\leq134$ and the extrapolated exponential  weight at $\tau=\tau_0$ between $6.0$ and $8.3\%$.  (This range is an operational estimate for the particular incoming state, propagation conditions and the mask parameters.)
This onset is consistent with the near-stationary  state to which the time
 propagation converges and which is illustrated by the independently normalized snapshot at
 $\tau=242$ in panel~(h) of 
Fig.~\ref{fig:Fig_Hussimi8snapshots}.

Thus the calculation provides a dynamical test of the stationary description:
  after the predominantly classical scattering components have escaped,
   evolution under the full finite-field relative motion Hamiltonian~\eqref{eq:Hr} leaves a
   central remnant whose density distribution, central zero, and decay rate are consistent
   with the independently calculated $m=1$ stationary resonance. Together, these
   observations support identifying the remnant with a metastable Laughlin-pair
   and give an operational estimate for the formation yield above $5\%$ for the
   particular collision considered here. This estimate is not a universal
   property of the resonance: its value depends on the incoming state and
   electrostatic setup,  while the operational definition of the yield also depends on the
   propagation and detection protocol used to distinguish the long-lived quasibound remnant
   from prompt Coulomb scattering. A systematic determination and optimization of the
   yield are beyond the scope of the present work. The propagation framework
   developed here provides a basis for such a study by allowing the incoming
   wave packet, electrostatic landscape, and readout criterion to be varied.
  

\section{Implications and outlook\label{sec:implicationsOutlook}}

The separation of time scales predicted here is already compatible with
depleted GaAs edge devices whose landscapes were not optimized for pair
transport.  Time-of-flight measurements give transverse confinement energies
$\hbar\omega_y=\SI{1.8}{meV}$ and $\SI{2.7}{meV}$ at $B=\SI{14}{T}$~\cite{Kataoka2016a}, while a separate
realization gives $\hbar\omega_y=\SI{2.7}{meV}$ at
$B=\SI{10}{T}$~\cite{Ubbelohde2023}.  At the latter field,
$\hbar/E_{\mathrm C}\simeq\SI{0.046}{ps}$.  
Modeling propagation along a locally straight edge by vanishing
longitudinal curvature $\kappa=0$, 
Eq.~\eqref{eq:mmax} maps these parameters to
$\bar m_{\mathrm{max}}\simeq6.0$, $3.5$, and $2.5$, respectively
($\chi\simeq0.0039$, $0.0088$, and $0.0147$).  In physical units, the
corresponding interaction times $t_I=(\hbar/E_{\mathrm C})\tau_I$ are
approximately $\SI{2.8}{ps}$, $\SI{1.3}{ps}$, and $\SI{0.9}{ps}$.  By comparison, the measured
drift velocities $v_{\mathrm d}\simeq(3$--$5)\times
10^4\,\SI{}{m/s}$ \cite{Kataoka2016a,Freise2019} imply flight times of order 
\SIrange{0.1}{0.3}{\nano\second}
over several micrometres.  Even the most conservative of these parameter sets
supports the odd $m=1$ branch. 
Using $\bar{m}_{\text{max}}=2.5$, the  numerical
  complex scaling gives the lifetime
  $t_1=(\hbar/E_{\mathrm C})(2\gamma_1)^{-1}\simeq\SI{46}{\nano\second}$.
  This is about six times shorter than the parabolic-saddle tunnelling estimate given by
  Eq.~\eqref{eq:gammam}, but remains more than two orders of magnitude longer
  than the device flight time.
  Thus,
within the present smooth-landscape model, existing depleted-edge device parameters
already support the ``long-lived'' regime required for propagation and leave
a substantial margin for future pair-specific optimization.

Existing electron-collision devices also reach the relevant formation regime.
At $B=\SI{10}{T}$, the symmetric-saddle parameters inferred by Ubbelohde
\emph{et al.} and used in the trajectory model of Fletcher \emph{et al.}
correspond to $\hbar\omega_y\simeq \SI{1.7}{meV}$ and $\SI{3.3}{meV}$,
respectively~\cite{Ubbelohde2023,Fletcher2023NatNa..18..727F}.  For
$\kappa=1$ these give
$\bar m_{\mathrm{max}}\simeq4.0$ ($\chi\simeq0.0058$,
$t_I\simeq\SI{1.6}{ps}$) and
$\bar m_{\mathrm{max}}\simeq1.6$ ($\chi\simeq0.0218$,
$t_I\simeq\SI{0.43}{ps}$).  Both landscapes therefore support the odd
$m=1$ resonance in the present model, while the softer saddle also supports
the $m=3$ branch.  These estimates locate the experimental devices on the
GaAs secondary axes of Figs.~\ref{fig:energiesCS1Dk1} and
\ref{fig:energiescomplexpartCS1Dk1}: the first figure gives the existence
threshold, and the second shows the rapid, exponential stabilization with
increasing $\bar m_{\mathrm{max}}$.  Their demonstrated field, confinement,
and timing scales therefore already overlap the regime in which such states
can be formed and remain intact during propagation.

These comparisons concern depleted devices in which the relevant dynamics
reduces to a few isolated electrons.  Pairing language is also
used for transport experiments on populated quantum Hall edges, where the
observables and microscopic objects are different.  Choi \emph{et al.}
measured conductance oscillations with flux period $h/(2e)$ and, in a
separate partition-noise measurement, inferred an effective partitioned
charge close to $2e$~\cite{Choi2015Pairing}.  Later interferometry showed that
this half-period contribution depends on the Landau-level character of the
adjacent edge channels~\cite{Biswas2023Pairing}.  In graphene, conductance gate
spectroscopy resolved frequencies equal to the sum of the Aharonov--Bohm
frequencies of two or three edge channels, interpreted as correlated phase
accumulation by electron pairs or triplets~\cite{Yang2024PairsTriplets}.  At
fractional filling factors $\nu$, Ghosh \emph{et al.} combined interference periods
$\Delta\Phi=h/(\nu e)$ with the distinct observation of a single-quantum-point-contact
shot-noise Fano factor $F=\nu$, and interpreted the two signatures together as
coherent bunching of two, three, or four elementary
anyons~\cite{Ghosh2025Bunching}.  Charging a central gate restored the
elementary-anyon interference period, which they described as dissociation,
although the Fano factor remained $F=\nu$~\cite{Ghosh2025Bunching}.  These
results establish correlated edge transport, but not necessarily a well-defined compound microscopic 
object: capacitive coupling can produce $h/2e$ oscillations
even when the interfering charge is
$e$~\cite{Werkmeister2024StronglyCoupled,Liang2025Capacitive}, while a noise
Fano factor characterizes the dominant partition process and can also depend
on neutral-mode dynamics~\cite{Ghosh2025Bunching}.  

Recent microscopic calculations address a different, bulk question: whether
fractional quasiparticles form an energetically bound molecule.  A collection
of anyons is a true molecule if its energy is lower when the anyons remain
together than when the same anyons are taken far apart.  For an ideal
two-dimensional Coulomb interaction, Gattu and Jain find such molecules at
$\nu=2/5$ and $3/7$, but not at
$\nu=1/3$~\cite{GattuJain2025MolecularAnyons}; Xu \emph{et al.} find that a
two-anyon molecule at $\nu=1/3$ becomes energetically favorable when the
electron interaction is sufficiently short-ranged~\cite{Xu2025AnyonClusters}.
More recent calculations show that nearby screening gates can likewise
stabilize $\nu=1/3$ anyon
molecules~\cite{Li2026BoundAnyons,WangZaletel2026AnyonMolecules}.  These go beyond a purely field-theoretic definition of anyons and 
emphasize that an anyon is an extended distortion of the surrounding quantum
Hall liquid rather than a point charge.  An anyon's internal density profile, the
range of the interaction, and the quantization of its relative motion
therefore determine whether binding occurs; fractional charge, statistics,
and fusion rules alone do
not~\cite{Li2026BoundAnyons,WangZaletel2026AnyonMolecules}.

The present work suggests how these bulk and edge questions could be related.
Isolated anyons also possess collective guiding-center coordinates, and their
mutual interaction can generate relative circulation rather than direct
radial separation; after projection onto the quasiparticle Hilbert space, this
motion is organized into discrete internal
orbitals~\cite{Xu2025AnyonClusters,Li2026BoundAnyons}.  Interaction can thus
produce a compact internal state, but its stability during transport is set by
the one-dimensional edge kinematics.  On a clean linear chiral branch, energy
and momentum conservation leave no spatially separating two-particle final
state: both decay products have the same group velocity, so neither can escape
the other.  A locally negative curvature, or negative effective mass, is still
more restrictive because positive internal energy cannot be converted into
relative kinetic energy within the same branch.  Weak positive curvature does
open a separating channel, but the relative velocity then vanishes with the
curvature and the required separation length can exceed the device length.
An internal anyon orbital may therefore lie \emph{above} the energy of
well-separated anyons and nevertheless remain intact throughout a transport
experiment.  Such a kinematically stabilized aggregate would be distinct both
from a below-threshold molecule and from correlated tunneling without a
compact internal state.  Real populated edges contain dispersion
nonlinearities, additional branches, disorder, and charge and neutral modes
that can reopen decay channels.  Whether they still leave a useful lifetime
window for a microscopic two-anyon state is the concrete question raised by
our results.

\section{Conclusions\label{sec:conclusions}}

We have characterized the stationary states, decay, and dynamical formation
of positive-energy Laughlin pairs in a smooth, locally depleted quantum Hall
channel.  The LLL stationary problem gives a curvature-controlled existence
threshold and number of quasibound branches, whose decay rates fall
exponentially as the states become better confined above the dissociation
threshold.  Their Husimi densities and wave-function zeros show how each
branch remains connected to an ideal Laughlin pair while acquiring outgoing
scattering structure.  Full finite-field propagation provides a complementary
dynamical test: after the prompt collision products escape, the remaining
two-electron core is consistent in structure and decay with the independently calculated
$m=1$ resonance.  Extrapolation of the late-time spatially uniform exponential decay gives an operational
formation-yield estimate above $5\%$ for the representative collision. Within the smooth-landscape model, mapping these results onto reported depleted-edge parameters gives lifetimes far longer than typical device flight
times and places existing propagation and collision devices in the relevant
stability and formation regimes.  This makes the controlled preparation,
transport, and detection of long-lived Laughlin pairs a realistic experimental
objective.

\acknowledgments

We thank Masaya Kataoka for discussions. 
This work has been  supported by the Latvian Council of Science (project no.\ lzp-2021/1-0232).  G.B.\ and V.K.\ 
has also been  supported by the project 23FUN05
AQuanTEC, which has received funding from the European Partnership on Metrology, co-financed from the European Union’s Horizon Europe Research and Innovation
Programme and by the Participating States.
The authors used \hyphenation{OpenAI} OpenAI Codex (gpt-5.6-sol) to assist with manuscript
  organization, technical editing,  and selected
  analytical and numerical checks. 
  All outputs were critically reviewed and independently verified by the authors, who take full responsibility for the
   manuscript's content.

\section*{Data Availability}

The numerical data and Mathematica notebooks used to reproduce the
figures in this article are publicly available from Ref.~\cite{dataset2026}.

\section*{Appendices}
\appendix
\section{Complex scaling method\label{app:Scaling} }

Laughlin pair states connected to the continuum by quadrupole perturbation become resonances -- non-normalizable exponentially decaying eigenstates  of time evolution, $\mathrm{H} \ket{\varepsilon_m} = (\varepsilon_m - \ii \gamma_m) \ket{\varepsilon_m}$.
 This eigenproblem can be solved by the complex scaling (CS) method proposed by Aguilar, Balslev, and Combes~\cite{aguilar1971class,balslev1971spectral}.  Rotating the coordinate from the real axis into the complex plane rotates the
  continuum spectrum, while the resonance eigenvalues exposed by the complex
  rotation remain unchanged and the corresponding eigenstates become localized
  and normalizable.
This transformation is represented by the dilation operator \cite{aguilar1971class}, $ S_\theta=\exp\left[+\theta(\hat{\mathrm{x}} \hat{\mathrm{
y}} 
+\hat{\mathrm{y}} \hat{\mathrm{
x}} )/2\right]$,  that rotates the coordinate-like $\hat{\mathrm{y}}$  and the momentum-like $\hat{\mathrm{x}}$ variables  in opposite directions by angle $\theta$, ${\left(\hat{\mathrm{x}}_\theta,\hat{\mathrm{y}}_\theta \right)=} {\left(S_\theta\hat{\mathrm{x}}S^{-1}_\theta,S_\theta\hat{\mathrm{y}}S^{-1}_\theta\right)=} {\left(\ee^{\ii\theta}\hat{\mathrm{x}},\ee^{-\ii\theta}\hat{\mathrm{y}}\right)}$, while
preserving their commutator.

Diagonalizing the dilated version of the Hamiltonian,
$\mathrm{H}_{\theta}=S_\theta \mathrm{H}S^{-1}_\theta
$, 
as a matrix in a suitably truncated Fock basis \eqref{eq:1eLaughlin} gives the desired eigenvalues and the rotated eigenfunctions,
$\mathrm{H}_{\theta} \ket{\varepsilon_{m} }_{\theta} = (\varepsilon_m - \ii \gamma_m) \ket{\varepsilon_{m} }_{\theta}$. The resonant state is calculated by rotating back, $\ket{\varepsilon_{m} } = S^{-1}_{\theta} \ket{\varepsilon_{m} }_{\theta}$.

The matrix elements $\braket{m|S_\theta|n}$ of the dilation operator
\cite{bargmannmatrixel1947,Varro_matrixel_2022}
in the Fock basis are zero
between $n$ and $m$ of mismatched parity, and equal to 
\begin{align} \label{eq:magicS}
\braket{m|S_\theta|n}
=
\frac{\sqrt{m!/n!}}{\left((m-n)/2\right)!} 
\left(\cos\theta\right)^{-\frac{m+n+1}{2}}\,
\left(\frac{-\ii}{2}\sin \theta \right)^{\frac{m-n}{2}}\, \nonumber\\
\times {}_2F_1\!\left(-\frac{n}{2},\,\frac{1-n}{2};\,\frac{m-n}{2}+1;\,\sin^2\theta\right)
\end{align}
for  $m\ge n$ and same parity $n$, $m$.
Here  ${}_2F_1$ is the
Gauss hypergeometric function. For $m <n$, complex conjugation of Eq.~\eqref{eq:magicS} provides the missing part of the transformation matrix,
$\braket{m|S_\theta|n}=\left(\braket{n|S_\theta|m}\right)^\ast$. As  $S_{\theta}$ is not unitary but complex orthogonal, $S^{-1}_{\theta}=S_{\theta}^{T}=S_{-\theta}$.

The matrix $\braket{m| \mathcal{H}|n}$ of the LLL-projected Hamiltonian \eqref{eq:projected} is parity-preserving and tridiagonal as $(\hat{\mathrm{x}}^2+\hat{\mathrm{y}}^2)$ is diagonal and 
\begin{align}\label{eq:quadrupolematrixelements}
  &  \braket{m|\left( \hat{\mathrm{x}}^2-\hat{\mathrm{y}}^2 \right) |n}= \nonumber\\
  & \sqrt{m(m-1)}\delta_{n,m-2}+ 
    \sqrt{(m+1)(m+2)}\delta_{n,m+2} \, .
\end{align}

In the present calculations, we use \(60\) basis functions, sufficient for
convergence over all parameter values shown in the figures at a fixed
complex-scaling angle $\theta=1/2$. Arbitrary-precision
arithmetic with a target precision of 140 decimal digits is used to ensure numerical stability  of exponentially small decay rates.

\section{Laughlin-pair resonances in a straight 1D dispersive edge\label{app:kappazero}}

Here we plot the analogues of the main-text Figures~\ref{fig:energiesCS1Dk1} and \ref{fig:energiescomplexpartCS1Dk1} for $\kappa=0$ which corresponds to the special case  of a straight 1D dispersive edge. The energies and the half-width of the odd-$m$ Laughlin-pair resonances are shown in Figures~\ref{fig:energiesCS1Dk0} and \ref{fig:gammask0}, respectively.
\begin{figure}
\includegraphics[width=0.499\textwidth]{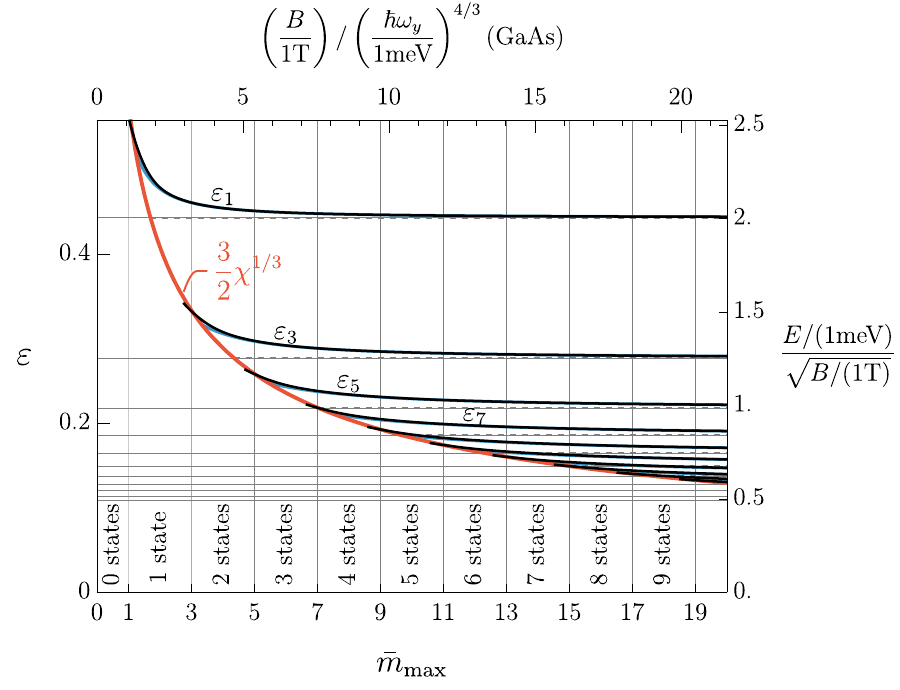}
 \caption{Dimensionless energies $\varepsilon_m=E_m/E_\text{C}$ of quasibound Laughlin pairs in a straight  waveguide with parabolic dispersion  ($\kappa=0$) in the large magnetic field limit (LLL approximation).  
    \label{fig:energiesCS1Dk0}}
\end{figure}

\begin{figure}
\includegraphics[width=0.48\textwidth]{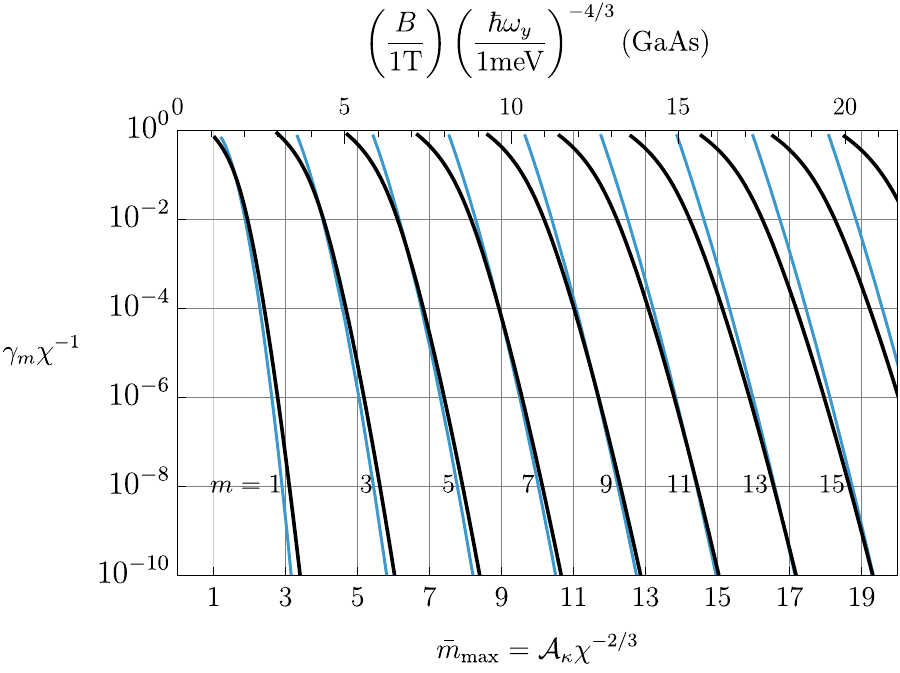}
\caption{Dimensionless resonance halfwidths, for  $\kappa=0$, scaled as
$\gamma_m/\chi$ plotted as
functions of the same parameter, $\bar{m}_{\text{max}}= \mathcal{A} _0\chi^{-2/3}$, as in Figure~\ref{fig:energiesCS1Dk0}.
\label{fig:gammask0}}
\end{figure}

\section{Wigner representation of resonance wavefunctions
\label{sec:WignerFunctions}}
Figure~\ref{fig:Wignerresonances} shows the Wigner functions defined in Eq.~\eqref{eq:WignerDef} for the same four resonance states and in the same panel order as Fig.\ref{fig:WFresonances}. The near-threshold states in panels (a)
and (c) exhibit sign-alternating fringes along the outgoing branches, whereas for the better-confined states in panels (b) and (d) the oscillatory
structure lies mainly inside the classical critical contour. Gaussian
smoothing of these patterns produces the simpler shape of the Husimi density in Fig.
 \ref{fig:WFresonances}. Because both resonance branches remain odd under
 inversion at finite $\chi$, evaluating Eq.\eqref{eq:WignerDef} at the origin
 gives $W(0)=-\pi^{-1}$ for a normalized state, saturating the lower bound,
  while the Wigner functions themselves are inversion symmetric, $W(z)=W(-z)$.

\begin{figure*}[htbp]
  \setlength{\tabcolsep}{2pt}
  \begin{tabular}{cc}
      \begin{overpic}[width=0.48\textwidth]{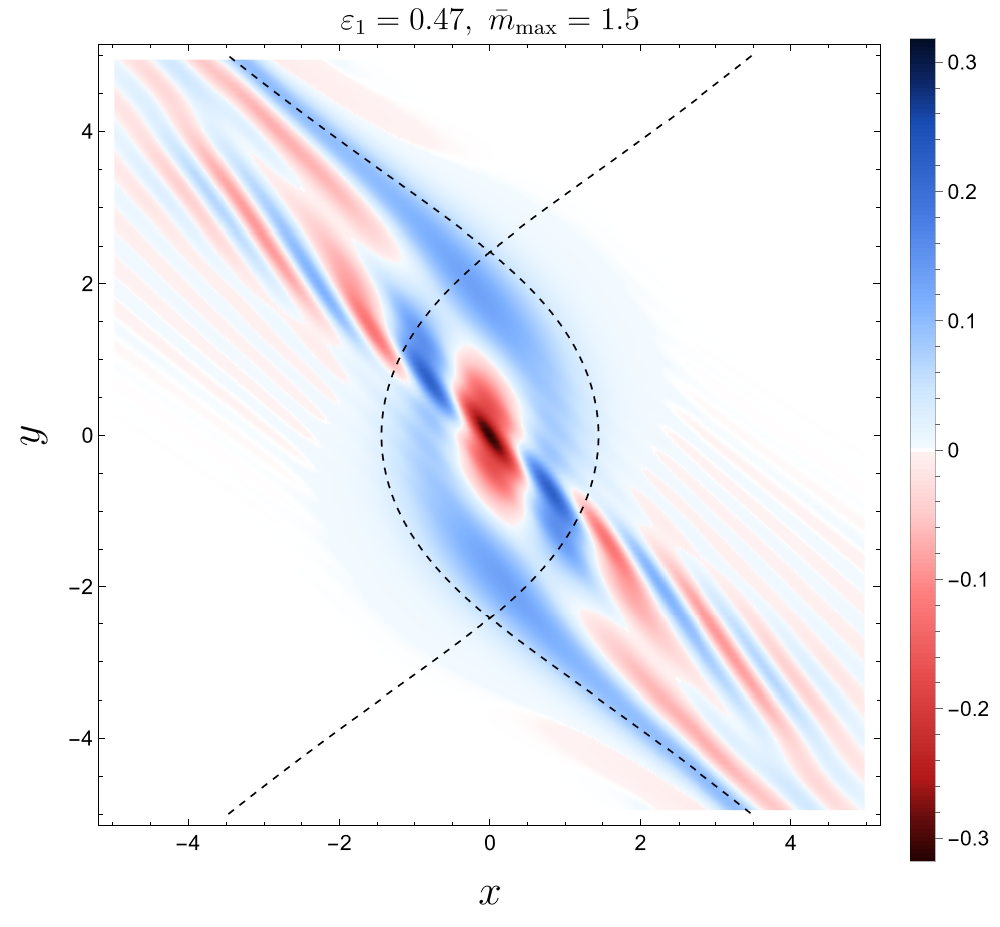}\put(2,90){\footnotesize\textbf{(a)}} \end{overpic} &
      \begin{overpic}[width=0.48\textwidth]{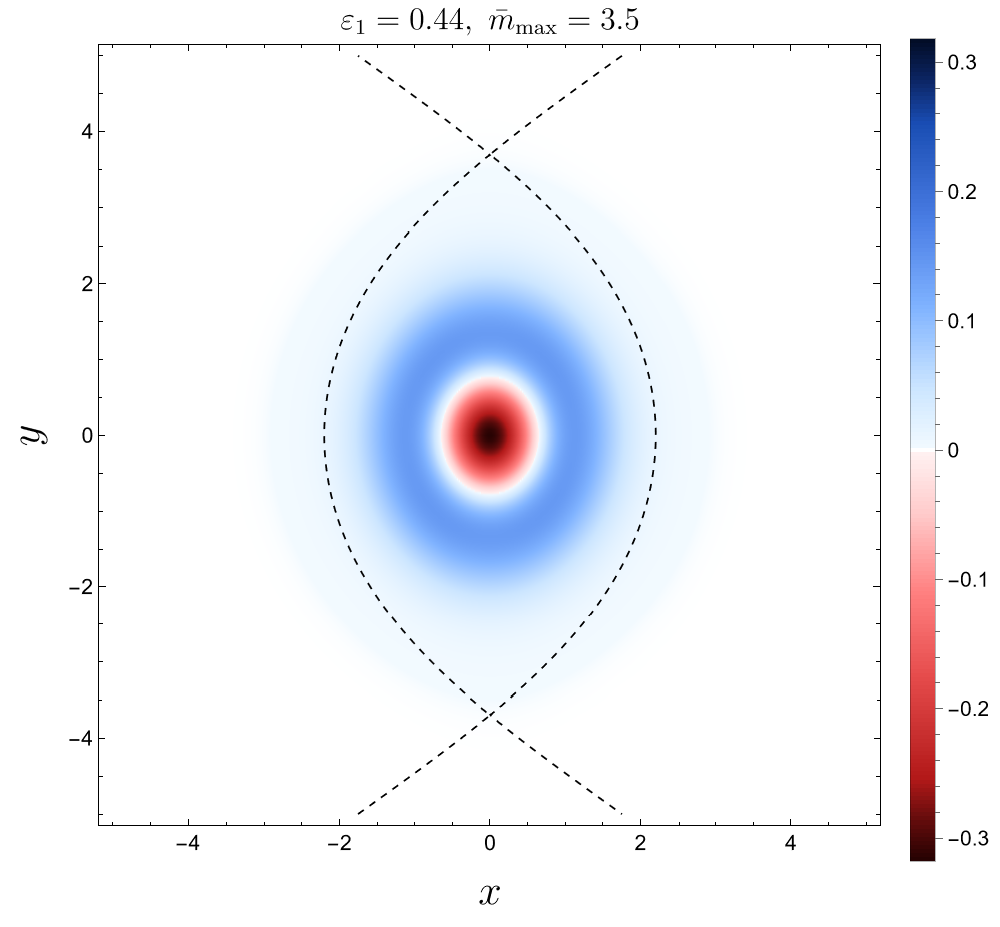}\put(2,90){\footnotesize\textbf{(b)}}\end{overpic} \\
      \begin{overpic}[width=0.48\textwidth]{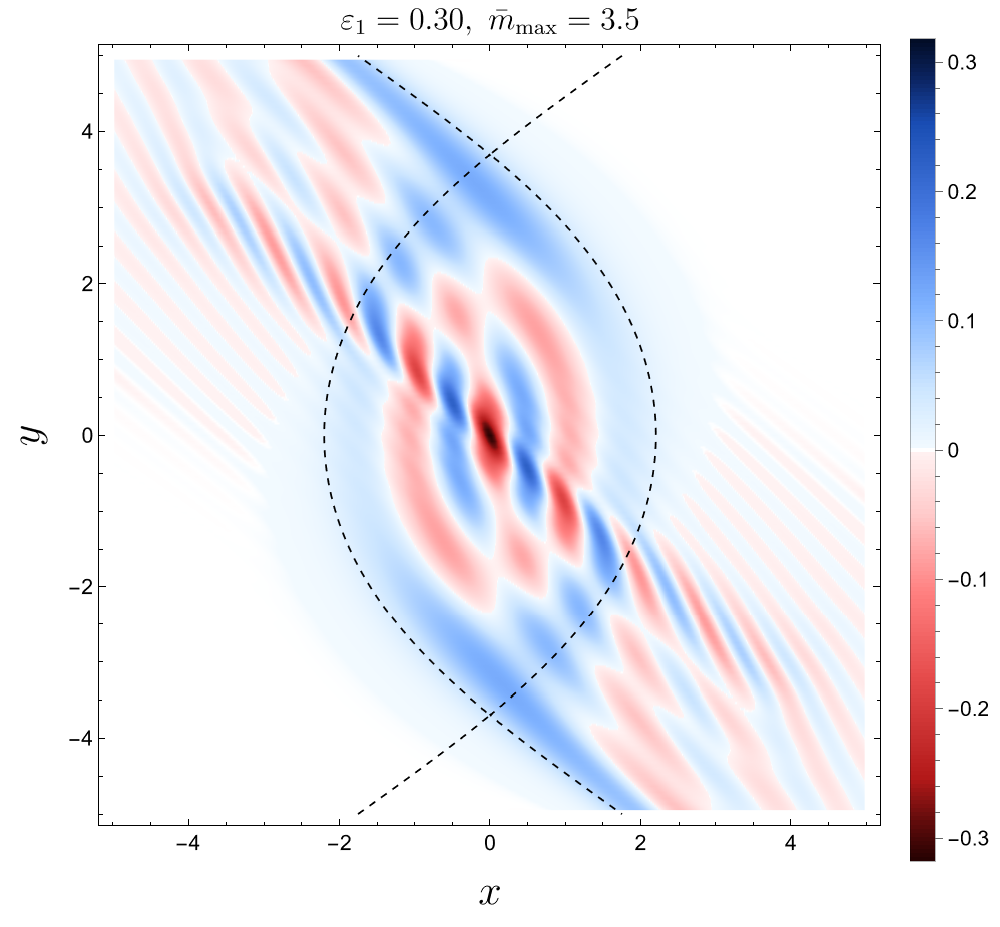}\put(2,90){\footnotesize\textbf{(c)}}\end{overpic} &
      \begin{overpic}[width=0.48\textwidth]{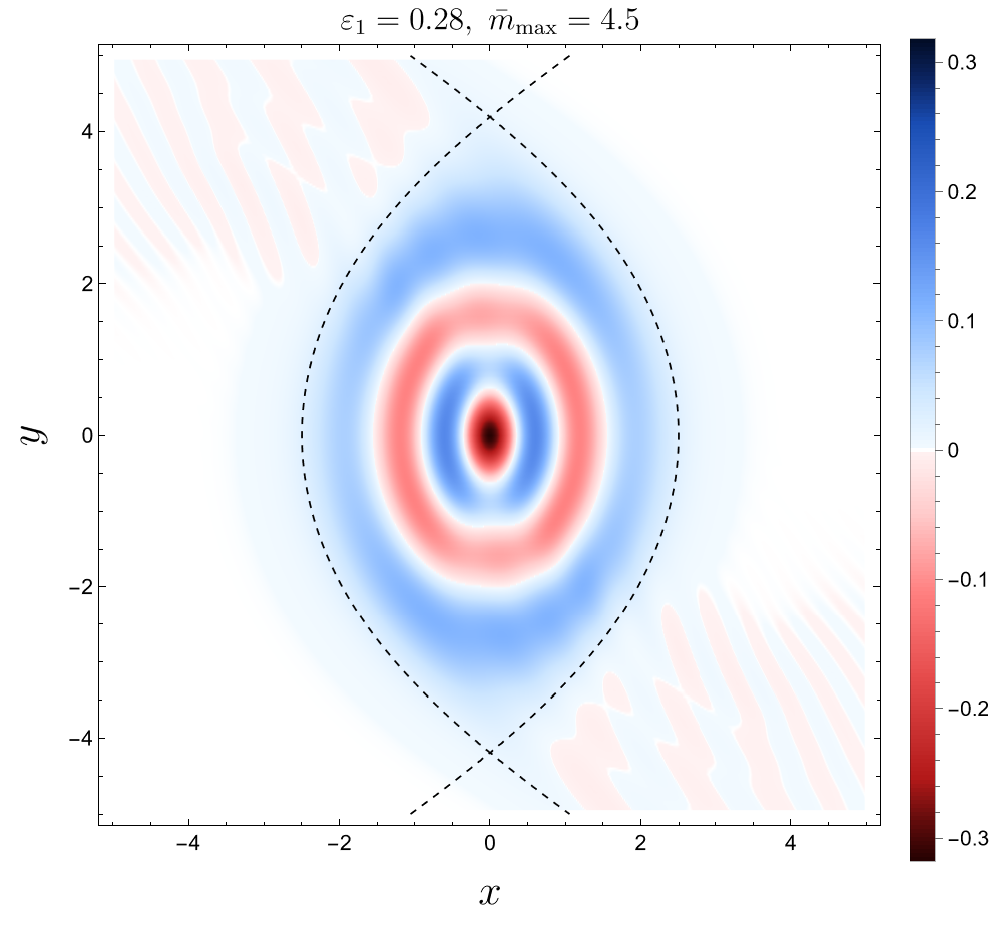}\put(2,90){\footnotesize\textbf{(d)}}\end{overpic}
  \end{tabular}
  \caption{Wigner functions $W(z)$ of Laughlin-pair resonances at  $\kappa=1$ (pure-quadrupole anisotropy). Parameter values are the same as in Figure~\ref{fig:WFresonances}: (a) and (b) show the $m=1$
  branch at $\bar m_{\mathrm{max}}=1.5$ and $3.5$, respectively; (c)
  and (d) show the $m=3$ branch at $\bar m_{\mathrm{max}}=3.5$ and $4.5$. The blue (red) color marks positive (negative) values of $W$.
  The corresponding resonance energies are printed above the panels.  The dashed curves
  are the classical critical level at
$\varepsilon=\varepsilon_{\mathrm{th}}$.\label{fig:Wignerresonances}}
\end{figure*}

\section{Implementation of time-dependent propagation\label{app:TDSE}}

\subsection{Initialization of the wave packet\label{app:TDSEintial}}

In the time-dependent calculation no LLL projection is made. The propagated
state is the full relative-coordinate wave function \(\psi(\bm r,t)\) governed
by \(\mathcal H_{\bm r}\), Eq.~\eqref{eq:Hr}. We use the Section~II units
\(l=\sqrt{2\hbar/(m_{\rm e}\omega_{\rm c})}\) and \(E_{\rm C}\); hence
\(x=r_x/l\) and \(y=r_y/l\) 
 are commuting grid coordinates 
in this Appendix.
Hatted variables \(\hat{\mathrm{x}}\) and \(\hat{\mathrm{y}}\) remain
reserved for LLL-projected guiding-center operators.

The initial state is specified at a large electron separation, where the
Coulomb term in \(\mathcal H_{\bm r}\) is negligible. The remaining quadratic
Hamiltonian is diagonalized by the Fertig-Halperin (FH) transformation
\cite{Fertig1987}, using the exact form described in Appendix~A.1 of
Ref.~\onlinecite{Pavlovska2023}. The transformed interaction-free Hamiltonian reads
\begin{align}
    \mathcal H_r^{(0)}
    =
    \frac{\hbar\omega_1}{2}(P^2-X^2)
    +
\frac{\hbar\omega_2}{2}(p^2+s^2) \, , 
\label{eq:FHdiagInit}
\end{align}
with canonical commutators $[X,P]=[s,p]=\ii$.
The pair \(X,P\) describes the unstable guiding-center motion, while \(s,p\)
describes quantized cyclotron modes. The ratio $\eta=\omega_1/ \omega_2$ controls the corresponding energy scales; the numerical initialization uses the
exact finite-$\eta$ transformation. Our  limit of interest is the large $B$-field limit where the Landau level separation dominates over the guiding potential effects. This corresponds to  $\eta \ll 1$.
Taking the $\eta \to 0$ limit 
makes the connection of FH transformation to the LLL projection transparent,
as for \(\eta\to0\) one has 
\(\omega_2\to\omega_{\rm c}\), \(\hbar\omega_1\to2\kappa\chi E_{\rm C}\), 
$X \to \sqrt{\kappa} \hat{x}$, and
$P \to -\hat{y}/\sqrt{\kappa} $
and
\begin{align}
    X=\sqrt\kappa\left(\frac{x}{2}-\ii\partial_y\right) \, ,
    \; 
    s=\frac{x}{2}+\ii\partial_y \, ,
    \label{eq:FHmixedStrongInit} \\
    P=-\frac{1}{\sqrt{\kappa}}\left(\frac{y}{2}+\ii\partial_x\right) 
   \, , \;
   p=\frac{y}{2}-\ii \partial_x \, .
   \label{eq:FHmixedStrongInit2}
\end{align}
The same limit identifies the FH guiding-center variables with the
LLL-projected coordinates of Sec.~II,
\begin{align}
    \hat{\mathrm{x}}=\frac{X}{\sqrt\kappa},
    \qquad
    \hat{\mathrm{y}}=-\sqrt\kappa\,P .
    \label{eq:FHtoLLLInit}
\end{align}
such that
$
    x =\hat{\mathrm{x}}+s$ and 
    $y=\hat{\mathrm{y}}+p $ and 
the commutation relations ensure  that $[r_x,r_y]=l^2 \, [x,y]=0$.
The guiding-center part of Eq.~\eqref{eq:FHdiagInit} then becomes the
quadratic term \(\chi(\hat{\mathrm{y}}^2-\kappa^2\hat{\mathrm{x}}^2)\) in
Eq.~\eqref{eq:projected} plus the   LLL offset $E_{\text{LL0}}$. 

The \(s\) oscillator is initialized in its ground state. In the strong-field
limit this is the cyclotron part of the LLL, while the
guiding-center part is a pure Gaussian,
\begin{align}
\psi_r(s,X)
=\frac{\ee^{-\frac{s^2}{2}}}{\sqrt[4]{\pi}}  
\frac{\ee^{-\frac{(X-\bar{X})^2}{4 \sigma_{\!  X}^2}
\left(1-2\ii\sigma_{\! X\!P} \right)
+\ii \bar{P} (X-\bar{X})}}{\sqrt[4]{2 \pi \sigma^2_{\!\! \scriptscriptstyle  X }}} \, .
\label{eq:gaussianWPsNew} 
\end{align}
The redundant momentum width is fixed by purity,
\begin{align}
    \sigma_P^2=\frac{1/4+\sigma_{XP}^2}{\sigma_X^2}.
    \label{eq:sigmaPpure}
\end{align}
The two real shape parameters \(\sigma_X\) and \(\sigma_{XP}\) therefore fix
both the probability-density ellipse and the quadratic phase of the incoming
Gaussian branch. The means \(\bar X,\bar P\) fix the branch center and its
guiding-center energy.

The shape parameters  $\sigma_X$ and $\sigma_{XP}$ can be translated into the covariance matrix $\Sigma$ of  $x$ and $y$ in  one Gaussian branch of the initial state.
Using the strong-field ($\eta \to 0$) expressions \eqref{eq:FHmixedStrongInit} and \eqref{eq:FHmixedStrongInit2}
\begin{align}
\Sigma_{xx} & = \frac{\sigma_X^2}{\kappa}+\frac12 \label{eq:sigmaxx} \\
\Sigma_{xy} =\Sigma_{yx} & = -\sigma_{XP} \\
\Sigma_{yy} & = \kappa\frac{1/4+\sigma_{XP}^2}{\sigma_X^2}
+\frac12 \label{eq:sigmayy}
\end{align}
The additional $1/2$ on the diagonal of $\Sigma$ comes from the  smearing by the uncertainty of the cyclotron motion.

The initial state in the relative coordinates in Eq.~\eqref{eq:gaussianWPsNew} is chosen so that it can
be completed to a two-electron state emitted by independent sources. For a pure
one-electron Gaussian written as 
\begin{align}
    \psi_j(\bm r_j)\propto
    \exp\left[
        -\frac14(\bm r_j-\bar{\bm r}_j)^{\intercal}
        \mathcal B_j
        (\bm r_j-\bar{\bm r}_j)
        +\frac{\ii}{\hbar}\bar{\bm p}_j\cdot\bm r_j
    \right],
    \label{eq:oneElectronComplexGaussian}
\end{align}
the real part of the symmetric complex matrix \(\mathcal B_j\) fixes the
inverse covariance matrix of \(|\psi_j|^2\), while its imaginary part fixes the
quadratic phase. Transforming to
\(\bm R=(\bm r_1+\bm r_2)/2\) and \(\bm r=\bm r_2-\bm r_1\), the only quadratic
term coupling \(\bm R\) and \(\bm r\) is proportional to
\(\bm R^{\intercal}(\mathcal B_2-\mathcal B_1)\bm r\). Thus, for this Gaussian
family, \(\psi_1(\bm r_1)\psi_2(\bm r_2)\) factorizes into center-of-mass and
relative-coordinate factors if and only if the two packets have the same full
Gaussian shape, \(\mathcal B_1=\mathcal B_2\). Their centers and mean momenta
are unconstrained by this condition. In the FH parametrization this amounts to
using the same \(\sigma_X,\sigma_{XP}\) and the same \(s\)-ground state in the
relative and center-of-mass sectors.
The covariance matrix $\Sigma$ is, up to the units, the covariance matrix of a single-electron Gaussian wave packet, $ \Sigma=2 \, l^{-2} (\operatorname{Re}\mathcal B_j)^{-1}$.
The center-of-mass factor is decoupled from
the Coulomb core and may be propagated analytically; only the relative factor is
needed in the TDSE calculation.

In the \(\eta\to0\) limit the wave function of the relative coordinate before antisymmetrization, Eq.~\eqref{eq:gaussianWPsNew}, becomes  
\begin{equation} 
   \varphi_{\mathrm G}(z)
=
 \varphi_{00}(z-\bar{z})
\exp\left[
\frac{\ii}{2}\operatorname{Im}\left(z\bar{z}^{*}\right)
\right] \,
, 
\label{eq:app:phizlimit}
\end{equation}
where $\bar z=\bar X/
    \sqrt{\kappa}-\ii \sqrt{\kappa}\bar P$. This is a product
of an anisotropic LLL state wave function~\cite{HaldaneAnisotropic2012}, 
\begin{equation} 
   \varphi_{00}(z)
=\frac{\sqrt[4]{1-|\lambda|^2}}{\sqrt{2\pi}}
\exp\left[ -\left(\frac{|z|^2}{4}-\frac{\lambda}{4}z^2\right)
\right]
\, 
,
\end{equation}
and the  magnetic translation phase factor in the symmetric gauge for the
shift of the position of  the center of the Gaussian from $0$ to $\bar{z}$. 
The anisotropy parameter
\begin{equation}
    \lambda=
    \frac{\Sigma_{xx}-\ii \Sigma_{xy}-1}{\Sigma_{xx}+\ii \Sigma_{xy}}
\end{equation} 
controls the amount of squeezing via its absolute value $|\lambda|$, while the phase $-\arg(\lambda)/2$ gives the angle of the major principal axis with respect to the $x$ axis.

The Gaussian in Eq. \eqref{eq:app:phizlimit} is a zero-free squeezed coherent state, but the exchange symmetry requirement   introduces zeros after antisymmetrization.
Electron exchange corresponds to changing the sign of the
relative coordinate \(\bm r\to-\bm r\), while leaving the center-of-mass coordinate unchanged.
Thus upon antisymmetrization  the LLL projection \eqref{eq:app:phizlimit} becomes 
\begin{equation} \label{eq:cat}
   \varphi_{\mathrm G}(z)-\varphi_{\mathrm G}(-z)
   \propto
   \varphi_{00}(z)
   \sinh\left[
      \frac{z}{2}
      \left(\bar{z}^{*}-\lambda \bar{z}\right)
   \right] .
\end{equation}

The zeros of the latter (and  hence of the corresponding Husimi density) are located at
\begin{equation} \label{eq:zerosLLL}
   z_n =
   \frac{2\pi \ii n}{\bar z^{*}-\lambda \bar z},
   \qquad n\in\mathbb{Z} \, .
\end{equation}
The zeros \eqref{eq:zerosLLL} are equally spaced on a straight line which is oriented perpendicular to the line connecting the complex plane origin with   $\bar z-(\lambda \bar z)^{*}$.
As long as the state of the relative coordinate remains close to the cat state Eq.~\eqref{eq:cat}, the spacing between adjacent zeros is asymptotically proportional to $|\bar{z}|^{-1}$.

In the calculation shown in Fig.~\ref{fig:Fig_Hussimi8snapshots} we use $\kappa=1$ and the initial state parameters 
\begin{align}
    \sigma_X=1/\sqrt{2} \, , \;  \sigma_{XP}=0 \, , \; 
    \bar X=-7 \, , \;  \bar P=8 \, ,
    \label{eq:TDSEinitParams}
\end{align}
which corresponds to LLL projection with  
    $\Sigma$ being the identity matrix and $\lambda=0$.
With $\chi=0.025$, the initial $\bar{z} = -7 -  8 \ii$ sets the (conserved) mean energy  of the incoming state $\bar{\varepsilon}= 0.441$, close to the threshold $\varepsilon_{\text{th}} =0.438$,  engaging the resonance at $\varepsilon_1 = 0.436$.

\subsection{Time propagation of the wave packet\label{app:TDSEpropagation} }

The time evolution of the wave packet in the relative coordinates is  calculated  numerically using the split-step Fourier method \cite{FeitFleck1982}. The  time evolution operator is factorized using the fourth-order symmetric Yoshida splitting \cite{YOSHIDA1990262}, which for $\Delta\tau\rightarrow0$  is given by
\begin{multline}
    \exp{\left(-\ii\Delta \tau\mathcal{H}_{\bm{r}}/E_\mathrm{C}\right)}=\mathcal{U}(a_1\Delta \tau) \, \mathcal{U}(a_2\Delta \tau) \, \mathcal{U}(a_1\Delta \tau) \, \\
    +O(\Delta \tau^5) \, ,
\end{multline}
where dimensionless time $\tau= t E_\mathrm{C}/\hbar$, and parameters $a_1=1/(2-2^{1/3})$ and $a_2=-2^{1/3}a_1$.
The second-order symmetric splitting of the sum of non-commuting operators is
\begin{align}
    \mathcal{U}(\Delta \tau)&=\ee^{-\ii \Delta \tau T_y/2}\ee^{-\ii \Delta \tau T_x/2}\ee^{-\ii \Delta \tau V} \ee^{-\ii \Delta \tau T_x/2}\ee^{-\ii \Delta \tau T_y/2} \, ,
\end{align}
where dimensionless kinetic energy operator $T_x$ is diagonal in the mixed coordinate-momentum representation of dimensionless relative coordinate $y$ and momentum space along $x$, and $T_y$ is diagonal in the mixed $x$ coordinate and $y$ momentum space for dimensionless relative coordinates:
 \begin{subequations}
 \begin{align}   
  T_x&=\frac{1}{2}\sqrt{\frac{\omega_\mathrm{c}}{\omega^{\star}}}\left(-\ii\partial_{x} +\frac{y}{2}\right)^2\\
    T_y&=\frac{1}{2}\sqrt{\frac{\omega_\mathrm{c}}{\omega^{\star}}}\left(-\ii\partial_{y} -\frac{x}{2}\right)^2 \, ,
\end{align}
\end{subequations}
and the dimensionless total potential energy in relative coordinates
\begin{align}
      V &=\chi(-\kappa^2x^2+y^2)+\frac{1}{\sqrt{2}\sqrt{x^2+y^2}} \, .
\end{align}
    The time-dependent propagation, Fig.~\ref{fig:Fig_Hussimi8snapshots}, is performed for a strong magnetic field ${\omega_\mathrm{c}/\omega^{\star}=25/4}$ and $\bar{m}_\mathrm{max}= \mathcal{A}_{\kappa} \chi^{-2/3}=1.5$, which corresponds to $\chi=0.025$ and $\eta=0.02$.
    
A Fermi-type mask function of the radial coordinate
\(r=\sqrt{x^2+y^2}\)
\begin{align}
    F(r)&=\left[1+\ee^{\left(r^4-r_\mathrm{  mask}^4\right)/L^4} \right]^{-1}
\end{align}
is applied to the wave function after each time step $\Delta \tau$ \cite{Kulander1992,FeitFleck1982} in order to absorb
the outgoing wave packet at the edge of the coordinate grid for $r \gtrsim r_\mathrm{  mask}$ outside the area shown in Fig.~\ref{fig:Fig_Hussimi8snapshots}. The parameter  $L$ controls the width of the corresponding transition region.
The wave function at the next time step is obtained as 
\begin{equation}
    \psi(\bm{r},\tau+\Delta \tau)=F(r)\exp{\left(-\ii\Delta \tau\mathcal{H}_{\bm{r}}/E_\mathrm{C}\right)}\psi(\bm{r},\tau) \, .
\end{equation}

In the calculation shown in Fig.~\ref{fig:Fig_Hussimi8snapshots}, the time step was \(\Delta\tau=2.5\cdot10^{-6}\).
The wave function was propagated on a rectangular
grid \(x\in[-70.70,70.56]\), \(y\in[-70.70,70.56]\) with
\(N_x\times N_y=2^{10}\times2^{10}\) grid points. The mask parameters were
\(r_\mathrm{  mask}=40.3\) and \(L=18.8\). The singularity of the Coulomb term at
\(r=0\) was regularized as  $1/r\rightarrow1/(r+a),~a=0.0071$. 

\bibliography{tdse-paper-biblio}

@article{Haldane1983,
  title = {{Fractional Quantization of the Hall Effect: A Hierarchy of Incompressible Quantum Fluid States}},
  author = {Haldane, F. D. M.},
  journal = {Phys. Rev. Lett.},
  volume = {51},
  issue = {7},
  pages = {605--608},
  numpages = {0},
  year = {1983},
  month = {Aug},
  publisher = {American Physical Society},
  doi = {10.1103/PhysRevLett.51.605},
  url = {https://link.aps.org/doi/10.1103/PhysRevLett.51.605}
}

@article{Varro_matrixel_2022,
doi = {10.1088/1367-2630/ac6b4d},
url = {https://dx.doi.org/10.1088/1367-2630/ac6b4d},
year = {2022},
month = {may},
publisher = {IOP Publishing},
volume = {24},
number = {5},
pages = {053035},
author = {Varró, Sándor},
title = {Coherent and incoherent superposition of transition matrix elements of the squeezing operator},
journal = {New Journal of Physics}
}

@article{bargmannmatrixel1947,
 ISSN = {0003486X, 19398980},
 URL = {http://www.jstor.org/stable/1969129},
 author = {V. Bargmann},
 journal = {Annals of Mathematics},
 number = {3},
 pages = {568--640},
 publisher = {[Annals of Mathematics, Trustees of Princeton University on Behalf of the Annals of Mathematics, Mathematics Department, Princeton University]},
 title = {{Irreducible Unitary Representations of the Lorentz Group}},
 urldate = {2025-09-19},
 volume = {48},
 year = {1947}
}

@article{girvin1984formalism,
  title={Formalism for the quantum {Hall} effect: {Hilbert} space of analytic functions},
  author={Girvin, SM and Jach, Terrence},
  journal={Physical Review B},
  volume={29},
  number={10},
  pages={5617},
  year={1984},
  publisher={APS}
}

@article{aguilar1971class,
  title={{A class of analytic perturbations for one-body Schr{\"o}dinger Hamiltonians}},
  author={Aguilar, Jacques and Combes, Jean-Michel},
  journal={Communications in Mathematical Physics},
  volume={22},
  pages={269--279},
  year={1971},
  publisher={Springer},
url = {https://doi.org/10.1007/BF01877510}
}

@article{balslev1971spectral,
  title={{Spectral properties of many-body Schr{\"o}dinger operators with dilatation-analytic interactions}},
  author={Balslev, Erik and Combes, Jean-Michel},
  journal={Communications in Mathematical Physics},
  volume={22},
  pages={280--294},
  year={1971},
  publisher={Springer},
  url = {https://doi.org/10.1007/BF01877511}
}

@article{Feve2007,
    title = {{An on-demand coherent single-electron source.}},
    year = {2007},
    journal = {Science (New York, N.Y.)},
    author = {F{\`{e}}ve, G and Mah{\'{e}}, A and Berroir, J.-M. and Kontos, T. and Pla{\c{c}}ais, B and Glattli, D. C. and Cavanna, A. and Etienne, B. and Jin, Y. and Feve, G. and Mahe, A. and Berroir, J.-M. and Kontos, T. and Placais, B. and Glattli, D. C. and Cavanna, A. and Etienne, B. and Jin, Y.},
    number = {5828},
    month = {5},
    pages = {1169--72},
    volume = {316},
    publisher = {American Association for the Advancement of Science},
    url = {https://science.sciencemag.org/content/316/5828/1169 https://science.sciencemag.org/content/316/5828/1169.abstract https://www.sciencemag.org/lookup/doi/10.1126/science.1141243 http://www.ncbi.nlm.nih.gov/pubmed/17525333},
    doi = {10.1126/science.1141243},
    issn = {1095-9203},
    pmid = {17525333}
}

@article{Laughlin1983,
    title = {{Anomalous Quantum Hall Effect: An Incompressible Quantum Fluid with Fractionally Charged Excitations}},
    year = {1983},
    journal = {Physical Review Letters},
    author = {Laughlin, R. B.},
    number = {18},
    month = {5},
    pages = {1395--1398},
    volume = {50},
    url = {https://link.aps.org/doi/10.1103/PhysRevLett.50.1395},
    doi = {10.1103/PhysRevLett.50.1395},
    issn = {0031-9007}
}

@article{Bocquillon2013,
    title = {{Coherence and indistinguishability of single electrons emitted by independent sources.}},
    year = {2013},
    journal = {Science (New York, N.Y.)},
    author = {Bocquillon, E and Freulon, V and Berroir, J-M and Degiovanni, P and Pla{\c{c}}ais, B and Cavanna, A and Jin, Y and F{\`{e}}ve, G},
    number = {6123},
    month = {3},
    pages = {1054--7},
    volume = {339},
    publisher = {American Association for the Advancement of Science},
    url = {http://science.sciencemag.org/content/339/6123/1054.abstract},
    doi = {10.1126/science.1232572},
    issn = {1095-9203},
    pmid = {23348504}
}

@article{Bauerle2018,
    title = {{Coherent control of single electrons: a review of current progress}},
    year = {2018},
    journal = {Reports on Progress in Physics},
    author = {B{\"{a}}uerle, Christopher and Christian Glattli, D and Meunier, Tristan and Portier, Fabien and Roche, Patrice and Roulleau, Preden and Takada, Shintaro and Waintal, Xavier},
    number = {5},
    month = {5},
    pages = {056503},
    volume = {81},
    publisher = {IOP Publishing},
    url = {http://stacks.iop.org/0034-4885/81/i=5/a=056503?key=crossref.442ccb56ea8ac1bb1f2f7dec6aa6bf74},
    doi = {10.1088/1361-6633/aaa98a},
    issn = {0034-4885}
}

@article{Fletcher2019,
    title = {{Continuous-variable tomography of solitary electrons}},
    year = {2019},
    journal = {Nature Communications},
    author = {Fletcher, J. D. and Johnson, N. and Locane, E. and See, P. and Griffiths, J. P. and Farrer, I. and Ritchie, D. A. and Brouwer, P. W. and Kashcheyevs, V. and Kataoka, M.},
    number = {1},
    month = {12},
    pages = {5298},
    volume = {10},
    url = {http://www.nature.com/articles/s41467-019-13222-1 http://arxiv.org/abs/1901.10985},
    doi = {10.1038/s41467-019-13222-1},
    issn = {2041-1723},
    arxivId = {1901.10985}
}

@article{Wang2022,
    title = {{Coulomb-mediated antibunching of an electron pair surfing on sound}},
    year = {2023},
    journal = {Nature Nanotechnology},
    author = {Wang, Junliang and Edlbauer, Hermann and Richard, Aymeric and Ota, Shunsuke and Park, Wanki and Shim, Jeongmin and Ludwig, Arne and Wieck, Andreas and Sim, Heung-Sun and Urdampilleta, Matias and Meunier, Tristan and Kodera, Tetsuo and Kaneko, Nobu-Hisa and Sellier, Hermann and Waintal, Xavier and Takada, Shintaro and B{\"{a}}uerle, Christopher},
    month = {10},
    volume={18},
    number={7},
    pages={721--726},
    url = {https://www.nature.com/articles/s41565-023-01368-5},
    arxivId = {2210.03452}
}

@article{Bocquillon2014,
    title = {{Electron quantum optics in ballistic chiral conductors}},
    year = {2014},
    journal = {Annalen der Physik},
    author = {Bocquillon, Erwann and Freulon, Vincent and Parmentier, François D. and Berroir, Jean-Marc and Pla{\c{c}}ais, Bernard and Wahl, Claire and Rech, Jérome and Jonckheere, Thibaut and Martin, Thierry and Grenier, Charles and Ferraro, Dario and Degiovanni, Pascal and F{\`{e}}ve, Gwendal},
    number = {1-2},
    month = {1},
    pages = {1--30},
    volume = {526},
    url = {http://doi.wiley.com/10.1002/andp.201300181},
    doi = {10.1002/andp.201300181},
    issn = {00033804}
}

@article{Bocquillon2012,
    title = {{Electron Quantum Optics: Partitioning Electrons One by One}},
    year = {2012},
    journal = {Physical Review Letters},
    author = {Bocquillon, E. and Parmentier, F. D. and Grenier, C. and Berroir, J.-M. and Degiovanni, P. and Glattli, D. C. and Pla{\c{c}}ais, B. and Cavanna, A. and Jin, Y. and F{\`{e}}ve, G.},
    number = {19},
    month = {5},
    pages = {196803},
    volume = {108},
    url = {https://link.aps.org/doi/10.1103/PhysRevLett.108.196803},
    doi = {10.1103/PhysRevLett.108.196803},
    issn = {0031-9007}
}

@article{Hermelin2011,
    title = {{Electrons surfing on a sound wave as a platform for quantum optics with flying electrons.}},
    shorttitle = {Nature},
    year = {2011},
    journal = {Nature},
    author = {Hermelin, Sylvain and Takada, Shintaro and Yamamoto, Michihisa and Tarucha, Seigo and Wieck, Andreas D and Saminadayar, Laurent and B{\"{a}}uerle, Christopher and Meunier, Tristan},
    number = {7365},
    month = {9},
    pages = {435--8},
    volume = {477},
    publisher = {Nature Publishing Group, a division of Macmillan Publishers Limited. All Rights Reserved.},
    url = {http://dx.doi.org/10.1038/nature10416},
    doi = {10.1038/nature10416},
    issn = {1476-4687},
    pmid = {21938064}
}

@article{Freulon2015,
    title = {{Hong-Ou-Mandel experiment for temporal investigation of single-electron fractionalization}},
    year = {2015},
    journal = {Nature Communications},
    author = {Freulon, V. and Marguerite, A. and Berroir, J.-M. and Pla{\c{c}}ais, B. and Cavanna, A. and Jin, Y. and F{\`{e}}ve, G.},
    number = {1},
    month = {11},
    pages = {6854},
    volume = {6},
    url = {http://www.nature.com/articles/ncomms7854},
    doi = {10.1038/ncomms7854},
    issn = {2041-1723}
}

@article{Dubois2013,
    title = {{Minimal-excitation states for electron quantum optics using levitons}},
    shorttitle = {Nature},
    year = {2013},
    journal = {Nature},
    author = {Dubois, J. and Jullien, T. and Portier, F. and Roche, P. and Cavanna, A. and Jin, Y. and Wegscheider, W. and Roulleau, P. and Glattli, D. C.},
    number = {7473},
    month = {10},
    pages = {659--663},
    volume = {502},
    publisher = {Nature Publishing Group, a division of Macmillan Publishers Limited. All Rights Reserved.},
    url = {http://dx.doi.org/10.1038/nature12713 http://www.nature.com/articles/nature12713},
    doi = {10.1038/nature12713},
    issn = {0028-0836},
    pmid = {24153178}
}

@article{Kulander1992,
  title = {Calculation of photoemission from atoms subject to intense laser fields},
  author = {Krause, Jeffrey L. and Schafer, Kenneth J. and Kulander, Kenneth C.},
  journal = {Phys. Rev. A},
  volume = {45},
  issue = {7},
  pages = {4998--5010},
  numpages = {0},
  year = {1992},
  month = {Apr},
  publisher = {American Physical Society},
  doi = {10.1103/PhysRevA.45.4998},
  url = {https://link.aps.org/doi/10.1103/PhysRevA.45.4998}
}

@article{Chaubaud2020,
  title = {Stellar Representation of Non-Gaussian Quantum States},
  author = {Chabaud, Ulysse and Markham, Damian and Grosshans, Fr\'ed\'eric},
  journal = {Phys. Rev. Lett.},
  volume = {124},
  issue = {6},
  pages = {063605},
  numpages = {6},
  year = {2020},
  month = {Feb},
  publisher = {American Physical Society},
  doi = {10.1103/PhysRevLett.124.063605},
  url = {https://link.aps.org/doi/10.1103/PhysRevLett.124.063605}
}

@article{McNeil2011,
    title = {{On-demand single-electron transfer between distant quantum dots}},
    year = {2011},
    journal = {Nature},
    author = {McNeil, R. P. G. and Kataoka, M. and Ford, C. J. B. and Barnes, C. H. W. and Anderson, D. and Jones, G. A. C. and Farrer, I. and Ritchie, D. A.},
    number = {7365},
    month = {9},
    pages = {439--442},
    volume = {477},
    url = {http://www.nature.com/articles/nature10444},
    doi = {10.1038/nature10444},
    issn = {0028-0836}
}

@article{Laughlin1982,
    title = {{Quantized motion of three two-dimensional electrons in a strong magnetic field}},
    year = {1983},
    journal = {Physical Review B},
    author = {Laughlin, R. B.},
    number = {6},
    month = {3},
    pages = {3383--3389},
    volume = {27},
    url = {https://link.aps.org/doi/10.1103/PhysRevB.27.3383},
    doi = {10.1103/PhysRevB.27.3383},
    issn = {0163-1829}
}

@article{Bisognin2019,
    title = {{Quantum tomography of electrical currents}},
    year = {2019},
    journal = {Nature Communications},
    author = {Bisognin, R. and Marguerite, A. and Roussel, B. and Kumar, M. and Cabart, C. and Chapdelaine, C. and Mohammad-Djafari, A. and Berroir, J.-M. and Bocquillon, E. and Pla{\c{c}}ais, B. and Cavanna, A. and Gennser, U. and Jin, Y. and Degiovanni, P. and F{\`{e}}ve, G.},
    number = {1},
    month = {12},
    pages = {3379},
    volume = {10},
    publisher = {Nature Publishing Group},
    url = {http://www.nature.com/articles/s41467-019-11369-5},
    doi = {10.1038/s41467-019-11369-5},
    issn = {2041-1723}
}

@misc{Silvestrov2025,
      title={{Theory of two-electrons optics experiments with smooth potentials: Flying electron molecules}}, 
      author={P. G. Silvestrov and Vyacheslavs Kashcheyevs and Patrik Recher},
      year={2025},
      eprint={2503.20325},
      archivePrefix={arXiv},
      primaryClass={cond-mat.mes-hall},
      url={https://arxiv.org/abs/2503.20325}, 
}

@article{Kataoka2016a,
    title = {{Time-of-Flight Measurements of Single-Electron Wave Packets in Quantum Hall Edge States}},
    year = {2016},
    journal = {Physical Review Letters},
    author = {Kataoka, M. and Johnson, N. and Emary, C. and See, P. and Griffiths, J. P. and Jones, G. A. C. and Farrer, I. and Ritchie, D. A. and Pepper, M. and Janssen, T. J. B. M.},
    number = {12},
    month = {3},
    pages = {126803},
    volume = {116},
    publisher = {American Physical Society},
    url = {http://journals.aps.org/prl/abstract/10.1103/PhysRevLett.116.126803 https://link.aps.org/doi/10.1103/PhysRevLett.116.126803},
    doi = {10.1103/PhysRevLett.116.126803},
    issn = {0031-9007}
}

@ARTICLE{Fletcher2023NatNa..18..727F,
       author = {{Fletcher}, J.~D. and {Park}, W. and {Ryu}, S. and {See}, P. and {Griffiths}, J.~P. and {Jones}, G.~A.~C. and {Farrer}, I. and {Ritchie}, D.~A. and {Sim}, H. -S. and {Kataoka}, M.},
        title = "{Time-resolved Coulomb collision of single electrons}",
      journal = {Nature Nanotechnology},
         year = 2023,
        month = jul,
       volume = {18},
       number = {7},
        pages = {727-732},
          doi = {10.1038/s41565-023-01369-4},
       adsurl = {https://ui.adsabs.harvard.edu/abs/2023NatNa..18..727F}
}

@article{Fertig1987,
    title = {{Transmission coefficient of an electron through a saddle-point potential in a magnetic field}},
    year = {1987},
    journal = {Physical Review B},
    author = {Fertig, H. A. and Halperin, B. I.},
    number = {15},
    month = {11},
    pages = {7969--7976},
    volume = {36},
    publisher = {American Physical Society},
    url = {https://link.aps.org/doi/10.1103/PhysRevB.36.7969},
    doi = {10.1103/PhysRevB.36.7969},
    issn = {0163-1829}
}

@article{Freise2019,
    title = {{Trapping and Counting Ballistic Nonequilibrium Electrons}},
    year = {2020},
    journal = {Physical Review Letters},
    author = {Freise, Lars and Gerster, Thomas and Reifert, David and Weimann, Thomas and Pierz, Klaus and Hohls, Frank and Ubbelohde, Niels},
    number = {12},
    month = {3},
    pages = {127701},
    volume = {124},
    url = {https://link.aps.org/doi/10.1103/PhysRevLett.124.127701},
    doi = {10.1103/PhysRevLett.124.127701},
    issn = {0031-9007}
}

@article{Ubbelohde2023,
    title = {{Two electrons interacting at a mesoscopic beam splitter}},
    year = {2023},
    volume= {18},
    pages = {733--740},
    journal = {Nature Nanotechnology},
    author = {Ubbelohde, Niels and Freise, Lars and Pavlovska, Elina and Silvestrov, Peter G. and Recher, Patrik and Kokainis, Martins and Barinovs, Girts and Hohls, Frank and Weimann, Thomas and Pierz, Klaus and Kashcheyevs, Vyacheslavs},
    url = {https://doi.org/10.1038/s41565-023-01370-x},
   }

@article{Pavlovska2023,
  title = {Collision of two interacting electrons on a mesoscopic beam splitter: Exact solution in the classical limit},
  author = {Pavlovska, Elina and Silvestrov, Peter G. and Recher, Patrik and Barinovs, Girts and Kashcheyevs, Vyacheslavs},
  journal = {Phys. Rev. B},
  volume = {107},
  issue = {16},
  pages = {165304},
  numpages = {14},
  year = {2023},
  month = {Apr},
  publisher = {American Physical Society},
  doi = {10.1103/PhysRevB.107.165304},
  url = {https://link.aps.org/doi/10.1103/PhysRevB.107.165304}
}

@article{Ferraro2013,
    title = {{Wigner function approach to single electron coherence in quantum Hall edge channels}},
    year = {2013},
    journal = {Physical Review B},
    author = {Ferraro, D. and Feller, A. and Ghibaudo, A. and Thibierge, E. and Bocquillon, E. and F{\`{e}}ve, G. and Grenier, Ch. and Degiovanni, P.},
    number = {20},
    month = {11},
    pages = {205303},
    volume = {88},
    publisher = {American Physical Society},
    url = {http://journals.aps.org/prb/abstract/10.1103/PhysRevB.88.205303},
    doi = {10.1103/PhysRevB.88.205303},
    issn = {1098-0121}
}

@article{Lutkenhaus1995NonclassicalSpace,
    title = {{Nonclassical effects in phase space}},
    year = {1995},
    journal = {Physical Review A},
    author = {L{\"{u}}tkenhaus, N. and Barnett, Stephen M.},
    number = {4},
    month = {4},
    pages = {3340--3342},
    volume = {51},
    publisher = {American Physical Society},
    url = {https://link.aps.org/doi/10.1103/PhysRevA.51.3340},
    doi = {10.1103/PhysRevA.51.3340},
    issn = {1050-2947}
}

@article{FeitFleck1982,
title = {{Solution of the Schrödinger equation by a spectral method}},
journal = {Journal of Computational Physics},
volume = {47},
number = {3},
pages = {412-433},
year = {1982},
issn = {0021-9991},
doi = {https://doi.org/10.1016/0021-9991(82)90091-2},
url = {https://www.sciencedirect.com/science/article/pii/0021999182900912},
author = {M.D Feit and J.A Fleck and A Steiger}
}

@article{YOSHIDA1990262,
title = {Construction of higher order symplectic integrators},
journal = {Physics Letters A},
volume = {150},
number = {5},
pages = {262-268},
year = {1990},
issn = {0375-9601},
doi = {https://doi.org/10.1016/0375-9601(90)90092-3},
url = {https://www.sciencedirect.com/science/article/pii/0375960190900923},
author = {Haruo Yoshida}
}

@article{Hudson1974WignerNonnegative,
  author  = {Hudson, R. L.},
  title   = {{When is the Wigner quasi-probability density non-negative?}},
  journal = {Reports on Mathematical Physics},
  volume  = {6},
  number  = {2},
  pages   = {249--252},
  year    = {1974},
  month   = oct,
  doi     = {10.1016/0034-4877(74)90007-X}
}

@article{CahillGlauber1969DensityOperators,
  author  = {Cahill, K. E. and Glauber, R. J.},
  title   = {{Density Operators and Quasiprobability Distributions}},
  journal = {Physical Review},
  volume  = {177},
  number  = {5},
  pages   = {1882--1902},
  year    = {1969},
  month   = jan,
  doi     = {10.1103/PhysRev.177.1882}
}

@article{HaldaneAnisotropic2012,
  title = {Model anisotropic quantum {Hall} states},
  author = {Qiu, R.-Z. and Haldane, F. D. M. and Wan, Xin and Yang, Kun and Yi, Su},
  journal = {Phys. Rev. B},
  volume = {85},
  issue = {11},
  pages = {115308},
  numpages = {9},
  year = {2012},
  month = {Mar},
  publisher = {American Physical Society},
  doi = {10.1103/PhysRevB.85.115308},
  url = {https://link.aps.org/doi/10.1103/PhysRevB.85.115308}
}

@article{complexzerossegura2013,
  title={Computing the complex zeros of special functions},
  author={Segura, Javier},
  journal={Numerische Mathematik},
  volume={124},
  number={4},
  pages={723--752},
  year={2013},
  publisher={Springer},
  url={https://link.springer.com/article/10.1007/s00211-013-0528-6}
}

@misc{video:tdsepropagation,
  note = {See Supplemental Material at [URL will be inserted by publisher to videotdse.mp4] for an animation showing the time dynamics of the probability density.}
}

@article{Yang2024PairsTriplets,
    title = {{Evidence for correlated electron pairs and triplets in quantum {H}all interferometers}},
    year = {2024},
    journal = {Nature Communications},
    author = {Yang, Wenmin and Perconte, David and D{\'{e}}prez, Corentin and Watanabe, Kenji and Taniguchi, Takashi and Dumont, Sylvain and Wagner, Edouard and Gay, Fr{\'{e}}d{\'{e}}ric and Safi, In{\`{e}}s and Sellier, Hermann and Sac{\'{e}}p{\'{e}}, Benjamin},
    number = {1},
    month = {11},
    pages = {10064},
    volume = {15},
    url = {https://doi.org/10.1038/s41467-024-54211-3},
    doi = {10.1038/s41467-024-54211-3},
    issn = {2041-1723}
}

@article{Ghosh2025Bunching,
    title = {{Coherent bunching of anyons and dissociation in an interference experiment}},
    year = {2025},
    journal = {Nature},
    author = {Ghosh, Bikash and Labendik, Maria and Umansky, Vladimir and Heiblum, Moty and Mross, David F.},
    number = {8069},
    month = {6},
    pages = {922--927},
    volume = {642},
    url = {https://doi.org/10.1038/s41586-025-09143-3},
    doi = {10.1038/s41586-025-09143-3},
    issn = {1476-4687}
}

@article{Choi2015Pairing,
    title = {{Robust electron pairing in the integer quantum {H}all effect regime}},
    year = {2015},
    journal = {Nature Communications},
    author = {Choi, H. K. and Sivan, I. and Rosenblatt, A. and Heiblum, M. and Umansky, V. and Mahalu, D.},
    number = {1},
    month = {6},
    pages = {7435},
    volume = {6},
    url = {https://doi.org/10.1038/ncomms8435},
    doi = {10.1038/ncomms8435},
    issn = {2041-1723}
}

@article{Shaju2025,
    doi = {10.1038/s41586-025-09139-z},
    year = {2025},
    pages = {928},
    title = {Evidence of {C}oulomb liquid phase in few-electron droplets},
    author= {Shaju, Jashwanth and Pavlovska, Elina and Suba, Ralfs and Wang, Junliang and Ouacel, Seddik and Vasselon, Thomas and Aluffi, Matteo and Mazzella, Lucas and Geffroy, Clement and Ludwig, Arne and Wieck, Andreas D and Urdampilleta, Matias Imanol and B{\"a}uerle, Christopher and Kashcheyevs, Vyacheslavs and Sellier, Hermann},
    volume = {642},
    journal = {Nature}}

@article{kaestner2010d,
    title = {{Generation of energy selective excitations in quantum Hall edge states}},
    year = {2011},
    journal = {Semiconductor Science and Technology},
    author = {Leicht, C and Mirovsky, P and Kaestner, B and Hohls, F and Kashcheyevs, V and Kurganova, E V and Zeitler, U and Weimann, T and Pierz, K and Schumacher, H W},
    number = {5},
    month = {5},
    pages = {055010},
    volume = {26},
    url = {http://arxiv.org/abs/1009.5897 http://stacks.iop.org/0268-1242/26/i=5/a=055010?key=crossref.106a08d8c4c3915cff5e1949e5f09c85},
    doi = {10.1088/0268-1242/26/5/055010},
    issn = {0268-1242}
}

@article{Pekola2013,
    title = {{Single-electron current sources: Toward a refined definition of the ampere}},
    shorttitle = {Rev. Mod. Phys.},
    year = {2013},
    journal = {Reviews of Modern Physics},
    author = {Pekola, Jukka P. and Saira, Olli-Pentti and Maisi, Ville F. and Kemppinen, Antti and M{\"{o}}tt{\"{o}}nen, Mikko and Pashkin, Yuri A. and Averin, Dmitri V.},
    number = {4},
    month = {10},
    pages = {1421--1472},
    volume = {85},
    publisher = {American Physical Society},
    doi = {10.1103/RevModPhys.85.1421},
    issn = {0034-6861}
}

@article{ROPP2015,
    title = {{Non-adiabatic quantized charge pumping with tunable-barrier quantum dots: a review of current progress}},
    year = {2015},
    journal = {Reports on Progress in Physics},
    author = {Kaestner, Bernd and Kashcheyevs, Vyacheslavs},
    number = {10},
    month = {10},
    pages = {103901},
    volume = {78},
    url = {https://iopscience.iop.org/article/10.1088/0034-4885/78/10/103901},
    doi = {10.1088/0034-4885/78/10/103901},
    issn = {0034-4885}
}

@article{Fletcher2013,
    title = {{Clock-Controlled Emission of Single-Electron Wave Packets in a Solid-State Circuit}},
    shorttitle = {Phys. Rev. Lett.},
    year = {2013},
    journal = {Physical Review Letters},
    author = {Fletcher, J. D. and See, P. and Howe, H. and Pepper, M. and Giblin, S. P. and Griffiths, J. P. and Jones, G. A. C. and Farrer, I. and Ritchie, D. A. and Janssen, T. J. B. M. and Kataoka, M.},
    number = {21},
    month = {11},
    pages = {216807},
    volume = {111},
    publisher = {American Physical Society},
    url = {http://link.aps.org/doi/10.1103/PhysRevLett.111.216807 https://link.aps.org/doi/10.1103/PhysRevLett.111.216807},
    doi = {10.1103/PhysRevLett.111.216807},
    issn = {0031-9007}
}

@article{Hillery1984DistributionFunctions,
  author  = {Hillery, M. and O'Connell, R. F. and Scully, M. O. and Wigner, E. P.},
  title   = {{Distribution functions in physics: Fundamentals}},
  journal = {Physics Reports},
  volume  = {106},
  number  = {3},
  pages   = {121--167},
  year    = {1984},
  doi     = {10.1016/0370-1573(84)90160-1},
  url     = {https://doi.org/10.1016/0370-1573(84)90160-1}
}

@article{Werkmeister2024StronglyCoupled,
  author = {Werkmeister, Thomas and Ehrets, James R. and Ronen, Yuval and
            Wesson, Marie E. and Najafabadi, Danial and Wei, Zezhu and
            Watanabe, Kenji and Taniguchi, Takashi and Feldman, D. E. and
            Halperin, Bertrand I. and Yacoby, Amir and Kim, Philip},
  title = {Strongly coupled edge states in a graphene quantum {Hall} interferometer},
  journal = {Nature Communications},
  volume = {15},
  pages = {6533},
  year = {2024},
  doi = {10.1038/s41467-024-50695-1}
}

@article{Biswas2023Pairing,
  author = {Biswas, Sourav and Kundu, Hemanta Kumar and Umansky, Vladimir and
            Heiblum, Moty},
  title = {{Electron Pairing of Interfering Interface-Based Edge Modes}},
  journal = {Physical Review Letters},
  volume = {131},
  pages = {096302},
  year = {2023},
  doi = {10.1103/PhysRevLett.131.096302},
  eprint = {2304.05140},
  archivePrefix = {arXiv}
}

@article{Liang2025Capacitive,
  author = {Liang, Shuang and Nakamura, James and Gardner, Geoffrey Charles and
            Manfra, Michael James},
  title = {Single electron interference and capacitive edge mode coupling
           generates {$\phi_0/2$} flux periodicity in {Fabry--P\'erot}
           interferometers},
  journal = {Nature Communications},
  volume = {16},
  pages = {7586},
  year = {2025},
  doi = {10.1038/s41467-025-62797-5}
}

@article{GattuJain2025MolecularAnyons,
  author = {Gattu, Mytraya and Jain, J. K.},
  title = {{Molecular Anyons in the Fractional Quantum Hall Effect}},
  journal = {Physical Review Letters},
  volume = {135},
  pages = {236601},
  year = {2025},
  doi = {10.1103/scl5-8pv6},
  eprint = {2505.22782},
  archivePrefix = {arXiv}
}

@article{Xu2025AnyonClusters,
  author = {Xu, Qianhui and Ji, Guangyue and Wang, Yuzhu and
            Trung, Ha Quang and Yang, Bo},
  title = {Dynamics of anyon clusters in fractional quantum {Hall} fluids},
  journal = {Physical Review B},
  volume = {112},
  pages = {235112},
  year = {2025},
  doi = {10.1103/vgz6-z98r},
  eprint = {2505.20257},
  archivePrefix = {arXiv}
}

@misc{Li2026BoundAnyons,
  author = {Li, Qingchen and Nosov, Pavel A. and Wang, Taige and Khalaf, Eslam},
  title = {Bound states of anyons: a geometric quantization approach},
  year = {2026},
  eprint = {2603.24701},
  archivePrefix = {arXiv},
  primaryClass = {cond-mat.str-el},
  note = {arXiv:2603.24701 [cond-mat.str-el]}
}

@misc{WangZaletel2026AnyonMolecules,
  author = {Wang, Taige and Zaletel, Michael P.},
  title = {Anyon molecules in fractional quantum {Hall} states},
  year = {2026},
  eprint = {2604.09798},
  archivePrefix = {arXiv},
  primaryClass = {cond-mat.mes-hall},
  note = {arXiv:2604.09798 [cond-mat.mes-hall]}
}

@dataset{dataset2026,
  author       = {Barinovs, Girts and
                  Buzs, Agris and
                  Kashcheyevs, Vyacheslavs},
  title        = {Dataset for publication "{Long-lived} {Laughlin} pairs
                   in a depleted quantum {Hall} edge channel"
                  },
  month        = sep,
  year         = 2026,
  publisher    = {Zenodo},
  doi          = {10.5281/zenodo.22816215},
  url          = {https://doi.org/10.5281/zenodo.22816215},
}

\end{document}